\documentclass[aps,prb,reprint,groupedaddress,showkeys]{revtex4-2}

\usepackage{amsmath}
\usepackage{amssymb}
\DeclareMathOperator{\sgn}{sgn} 
\DeclareMathOperator{\im}{Im} 
\DeclareMathOperator{\trace}{Tr}
\newcommand{\dd}{\text{d}} 
\newcommand{\ii}{\text{i}} 

\usepackage{graphicx}

\usepackage[hidelinks]{hyperref}

\begin{document}

\title{Parquet approximation and one-loop renormalization group: Equivalence on the leading-logarithmic level}

\author{Jan Diekmann}
\author{Severin G. Jakobs}

\affiliation{Institut f\"ur Theorie der Statistischen Physik, RWTH Aachen University, 52056 Aachen, Germany\\
and JARA---Fundamentals of Future Information Technology, 52056 Aachen, Germany}

\date{\today}

\begin{abstract}
We investigate the functional renormalization group (FRG) flow of the two-particle vertex function of a model for X-ray absorption in metals.
Concerning the appearance of logarithmic divergences, the model is prototypical for an important class of mostly zero- and one-dimensional systems which includes the Kondo model and the interacting one-dimensional Fermi gas.
For our analysis we formulate the FRG in the framework of the real-time zero-temperature formalism, in which the model was studied before with a parquet-based approach.
We establish that a reasonably crafted, purely-fermionic one-loop FRG approximation is fully equivalent on a detailed level to the leading-logarithmic parquet approximation.
These two approximation schemes are thus found to just represent different perspectives on the same technical steps.
This finding also reconfirms the traditional understanding of the capabilities of one-loop RG approximations for such models, which was recently put into question by an investigation of the X-ray-absorption model with multiloop FRG.
\end{abstract}

\maketitle

\section{Introduction}

It is a well-known problem that perturbative approximations to low-dimensional condensed-matter systems of interacting fermions lead to expressions that diverge logarithmically at low energies.
The focus of this paper lies on a class of mostly zero- and one-dimensional systems characterized by the simple pattern in which logarithmic divergences arise in particle-particle and particle-hole bubbles of perturbative diagrams for two-particle correlation functions.
This class includes a model for X-ray absorption in metals \cite{Mahan67}, the Kondo model \cite{Abrikosov65}, and the Fermi gas model for one-dimensional conductors \cite{Solyom79}.
Technically, the divergences appear when the integral of a Green-function resolvent with respect to single-particle energy is cut off by the Fermi edge of the level occupancy.
Physically, this phenomenon is reflected by the parameter dependence of a related susceptibility.
This can be given, e.g., by a power law with an exponent that depends on the interaction: an expansion in powers of the interaction then leads to the logarithmic contributions.

The leading-logarithmic parquet approximation provides a means to compute the correlation function in the vicinity of the divergence.
The underlying reasoning is as follows.
The divergent terms resulting from perturbative diagrams of different order and structure depend on different powers of the interaction and of the logarithms.
For small interaction and not too close to the divergence of the logarithms, the diagrams can be grouped into leading contributions and negligible corrections.
Summing the ``leading logarithms'' then yields a controlled approximation in this regime.
For the class of models which we focus on, all leading contributions are contained in the so-called parquet diagrams (with bare lines), which comprise the ladder diagrams in the particle-particle and particle-hole channels and also diagrams that result from crossing the channels.
The strategy to derive the leading-logarithmic approximation from the leading contributions of all parquet diagrams was first implemented for meson scattering \cite{Diatlov57}.
Well-known realizations of this concept for low-dimensional condensed-matter systems include the application to the problem of X-ray absorption in metals \cite{Roulet69}, to the Kondo problem \cite{Abrikosov65}, and to fermions in one dimension with a short-ranged interaction \cite{Bychkov66}.
In the last-mentioned application, the leading-logarithmic approximation seems to wrongly predict a finite-temperature phase transition for attractive interaction.
This is, however, beyond its regime of applicability: at low temperatures the neglected lower-order logarithmic contributions become important \cite{Solyom79}.

Apart from the \emph{leading-logarithmic} version of the parquet approximation, there exists another well-established one \cite{DeDominicis64, Bickers91}, which we refer to as the \emph{full} parquet approximation.
While both versions are based on parquet diagrams, they differ in technical aspects, in the kind of system they are typically applied to, and in the justification for their use.
The technical considerations of this paper concern specifically the leading-logarithmic parquet approximation.
However, in the next paragraph we briefly describe the full parquet approximation since a clear understanding of the differences of both approaches will become necessary to classify our findings and relate them to the multiloop functional renormalization group (multiloop FRG).

In our class of mostly zero- and one-dimensional models, each propagator bubble in one of the relevant channels produces a simple logarithmic divergence.
The situation is more complicated for many two-dimensional models of correlated fermions: depending on the system parameters, on the filling, and on the channel under consideration, the bubbles of two-dimensional models can feature either no divergence or a logarithmic or a squared logarithmic one, see, e.g., Ref.~\cite{Irkhin01, Katanin09}.
It is then more involved to identify the leading contributions. Furthermore, in order to decide on the existence and location of phase transitions, subleading contributions might be relevant.
Nonetheless, summing up all parquet diagrams is a well-known approximation strategy for two-dimensional problems \cite{DeDominicis64, Bickers91}.
The typically applied full parquet approximation takes into account all particle-hole channels, uses propagator lines that are dressed with a self-energy determined self-consistently from a Schwinger-Dyson equation, and evaluates the exact sum of all parquet diagrams.
In these aspects it differs from the leading-logarithmic parquet approximation for systems with simple logarithmic divergences, which takes into account only the leading-logarithmic particle-hole channel, uses bare propagator lines, and evaluates only the leading-logarithmic part of the corresponding parquet diagrams.
While in the leading-logarithmic parquet approximation the totally irreducible vertex is replaced by the bare one, there exist extensions of the full parquet approximation which use more involved approximations for the totally irreducible vertex: in the parquet dynamical vertex approximation \cite{Toschi07}, e.g., it is approximated by the local vertex resulting from dynamical mean-field theory.
Due to the complicated logarithmic structure, the full parquet approximation for two-dimensional systems is usually not known to be controlled.
It is still considered to be beneficial as it includes fluctuations in different channels of pair propagation in an unbiased way, respects the crossing symmetry \cite{Bickers91} and related sum rules \cite{Vilk97}, satisfies one-particle conservation laws \cite{Kugler18c}, and is understood to comply with the Mermin-Wagner theorem \cite{Bickers92}.
Such arguments may also motivate the use of the full parquet approximation for other than two-dimensional models.
For example, Ref.~\cite{Pudleiner19} shows an application of this approximation to a model for a benzene molecule.
When the full parquet approximation is applied to models for which the leading-logarithmic parquet approximation is controlled, both approximations coincide on the leading-logarithmic level.

In this paper we examine the relation between the leading-logarithmic parquet approximation and a specific renormalization group (RG) approximation.
Indeed, scaling arguments and the RG provide another approach to interacting fermionic systems in low dimensions.
Historically, the development of these techniques for zero- and one-dimensional systems was driven by the quest for approximations beyond the parquet-based leading-logarithmic one \cite{Gruner74, Hewson93, Solyom79}.
In early approaches field-theoretical RG techniques were applied to the Kondo
problem \cite{Abrikosov70, Fowler71}, to the weakly interacting one-dimensional Fermi gas \cite{Kimura73, Solyom79} (see as well \cite{Shankar94}), and also to the problem of X-ray absorption in metals \cite{Solyom74}.
In all these applications, the lowest-order approximation reproduced the leading-logarithmic result known from the respective parquet treatments \cite{Abrikosov65, Roulet69, Bychkov66}.
Also Anderson's ``poor man's scaling'' approach to the Kondo problem \cite{Anderson70, Solyom74a} and its application to fermions in one dimension \cite{Solyom79} reproduce the respective leading-logarithmic results.
For a specific two-dimensional model, the equivalence of a one-loop RG and a parquet approximation on the leading-logarithmic level is discussed in Ref.~\cite{Binz03}; the model has a nested Fermi surface but no van Hove singularity such that the bubbles produce simple logarithmic divergences only.
For general two-dimensional systems, an equivalence of one-loop RG and parquet approaches is not expected.
For our class of mostly zero- and one-dimensional models, however, the RG idea leads beyond the leading logarithms.
A cornerstone is the accurate description of the Kondo effect by Wilson's numerical RG \cite{Wilson75}.
But even an RG flow that is constructed just to account for the leading logarithms can lead to predictions beyond the realm of parquet approximations if it is understood as connecting models with identical low-energy properties.
The underlying concept of universality classes shaped today's understanding of one-dimensional interacting fermionic systems as being Luttinger liquids, Luther-Emery
liquids, or Mott insulators \cite{Solyom79, Voit95}.

Recently, the relation between the parquet approximations and the RG, now in form of the FRG \cite{Metzner12, Kopietz10, Platt13}, came again into focus, leading to the construction of the so-called multiloop FRG.
Starting point of that development was an FRG study of X-ray absorption in metals by Lange et al. \cite{Lange15}.
Using Hubbard-Stratonovich transformations and a low-order truncation scheme, Lange et al. reproduced the result of the leading-logarithmic parquet approximation \cite{Roulet69} for the X-ray response function.
Kugler and von Delft scrutinized this FRG approach and came to the conclusion that its success is fortuitous \cite{Kugler18}.
Among other criticism Kugler and von Delft point out that the scheme of Lange et al. only reproduces ladder diagrams.
Such a diagram-wise juxtaposition of FRG and parquet approximations is possible as the FRG flow can be interpreted on the level of individual, flowing diagrams \cite{Jakobs07}.
Kugler and von Delft expanded this idea rigorously and constructed a multiloop extension to a purely-fermionic one-loop FRG which makes it possible to approach the exact sum of all parquet diagrams via iterative one-loop extensions.
They provided schemes to approach the sum of parquet diagrams with either bare lines \cite{Kugler18a} or self-consistently dressed lines \cite{Kugler18b} and for different approximations for the totally irreducible vertex \cite{Kugler18c}.

As the multiloop FRG offers the possibility to compute self-consistent parquet-based approximations by solving flow equations, it is seen as a promising tool for the study of correlated two-dimensional systems.
Recently, it was combined with special approaches to the momentum dependence and high-frequency asymptotics of the two-particle vertex and applied to the two-dimensional Hubbard model \cite{Tagliavini19, Hille20a, Hille20}.
This allowed to reduce the pseudo-critical temperature of antiferromagnetic ordering compared to one-loop FRG \cite{Tagliavini19}, to achieve numerical convergence to results of the full parquet approximation and of determinant quantum Monte Carlo up to moderate interaction strength \cite{Hille20a}, and to analyze pseudo-gap physics at weak coupling \cite{Hille20}.
Including multiloop corrections could also be beneficial for the RG study of two-dimensional quantum spin systems (in pseudo-fermion representation) because a two-loop extension was already found to attenuate the violation of the Mermin Wagner theorem \cite{Rueck18}.
Furthermore, the multiloop scheme based on an irreducible vertex from dynamical mean-field theory could provide a viable alternative way to evaluate the parquet dynamical vertex approximation \cite{Rohringer18, Kugler18c}.

Whereas these applications concern the full parquet approximation for two-dimensional systems, Kugler and von Delft motivated and introduced the multiloop scheme in the context of X-ray absorption in metals \cite{Kugler18, Kugler18a}.
The interacting region in the corresponding model is zero-dimensional and the propagator bubbles in the two relevant channels produce simple logarithmic divergences.
Due to its basic structure, this model can in fact be seen as prototypical for the case that the parquet diagrams with bare lines comprise the leading logarithmic contributions.
Correspondingly, Nozi\`eres and collaborators understood their parquet study of this model as a preparation for the analysis of more complicated models with that structure of logarithms like the Kondo model \cite{Roulet69, Nozieres69}.
Formulated from the RG perspective, the model for X-ray absorption in metals is at the core of the class of (mostly zero- and one-dimensional) models for which a reasonably crafted lowest-order, i.e., one-loop, RG approximation is understood to be accurate and equivalent to the parquet approximations on the leading-logarithmic level.
This conventional conception is in surprising contrast to what Kugler and von Delft report from their multiloop FRG study of that model \cite{Kugler18a} -- namely that increasing the number of loops improves the numerical results with respect to the known solution of Nozi\`eres and collaborators \cite{Roulet69, Nozieres69, Nozieres69a}.
In an astounding twist, the multiloop approach of Ref.~\cite{Kugler18a} inverts the direction of the historical development of techniques for such models: while the RG was originally used as a conceptual framework to overcome the restrictions inherent to the leading-logarithmic parquet approximation, Ref.~\cite{Kugler18a} transforms its modern functional formulation into a tool to evaluate the exact sum of the parquet diagrams (in this case with bare lines and without the subleading particle-hole channel).
For the model under consideration, this sum constitutes no controlled improvement compared to the leading-logarithmic parquet approximation.
It differs subleadingly without being exact on a subleading level \cite{Nozieres69}.
Since Ref.~\cite{Kugler18a}, however, reports on improvements compared to one-loop FRG, there emerges the pressing question whether the latter might be deficient: is a one-loop FRG approximation without multiloop extensions really less accurate than the early implementations of RG and poor man's scaling that were already able to reproduce the leading-logarithmic parquet results?

In this paper we establish that for the considered class of systems with simple logarithmic divergences a reasonably crafted one-loop FRG approximation is fully equivalent to the leading-logarithmic parquet approximation.
We do so by constructing, specifically for the problem of X-ray absorption in metals, a one-loop FRG approximation that is in fact identical on a detailed level to the leading-logarithmic approximation procedure which was performed by Roulet et al. within the parquet formalism \cite{Roulet69}.
The only formal difference is that the cutoff is introduced at a different stage of the derivation without influencing the result.
From this viewpoint the two approaches actually fuse into one.

In order to allow for a detailed comparison to the leading-logarithmic parquet approximation of Roulet et al.~\cite{Roulet69}, we devise our FRG approximation in the framework of the real-time zero-temperature formalism, also known as ground-state formalism.
For brevity we refer to this formalism simply as the zero-temperature formalism.
As the FRG for condensed-matter systems was so far used within the Matsubara and the Keldysh formalism \cite{Metzner12}, we first need to transfer the formulation of the method to the zero-temperature formalism.
Our approach to that formalism is inspired by Ref.~\cite{Negele88} but differs in some respects.
In particular, we develop a functional-integral representation of the generating functional that is based on standard coherent states and is therefore easily applicable to the interacting case.
Then we perform the usual steps to derive the flow equations for the one-particle irreducible (1PI) vertex functions.

The paper is organized as follows.
We briefly introduce the model under investigation in Sec.~\ref{sec:model}.
The most important features of a perturbative approach to it are recapped in Sec.~\ref{sec:perturbation_theory}.
In particular, the occurrence of logarithmic divergences is discussed.
In Sec.~\ref{sec:FRG_general_models} we set up the FRG framework within the zero-temperature formalism for a general model.
Some details on deriving the diagrammatic expansion and the flow equations are relocated from this section to the Appendix.
The core of the paper is Sec.~\ref{sec:FRG_approach}, where we construct our one-loop FRG approximation and establish its full equivalence to the leading-logarithmic parquet approximation of Ref.~\cite{Roulet69}.
Finally, Sec.~\ref{sec:conclusion} provides a conclusion and outlook.

\section{Model}
\label{sec:model}

In this section we briefly introduce the model under consideration.
It is essentially taken from Ref.~\cite{Roulet69}, where more details can be found.

The investigated basic model provides a description of the X-ray-absorption singularity in metals.
It comprises two electronic bands: the conduction band and some lower-energy band.
The latter is assumed to be flat as it typically originates from atomic orbitals that are more localized.
As such it can be represented by a single so-called deep state.
The effect of intraband Coulomb interaction, which leads to long-lasting quasi-particle states, is assumed to be already accounted for by effective single-particle parameters.
The interaction that is considered explicitly is an attractive one between the conduction electrons and a hole at the deep state.
The electron spin is neglected.
This physical model is described by the Hamiltonian
\begin{equation}
\label{eq:Hamiltonian}
H = \sum_k \varepsilon_k a_k^\dagger a_k + \varepsilon_d a_d^\dagger a_d - \frac{U}{V} \sum_{kk'} a_{k'}^\dagger a_k a_d a_d^\dagger.
\end{equation}
Here, $a_d^\dagger$ creates an electron in the deep state and $a_k^\dagger$ creates an electron with momentum $k$ and energy $\varepsilon_k$ in the conduction band, which shall have a constant density of states $\rho$ and a bandwidth $2 \xi_0$.
We set the zero of single-particle energy in the middle of the conduction band such that $\varepsilon_k \in [- \xi_0, \xi_0]$.
Then the deep-state eigenenergy is $\varepsilon_d < - \xi_0$.
The interaction amplitude $U > 0$ is assumed to be momentum independent and thus describes a local interaction in real space.
$V$ denotes the volume.
We study the system at vanishing temperature $T=0$ and with a half-filled conduction band, i.e., the Fermi energy is $\varepsilon_F = 0$.
In the resulting ground state, the deep level is occupied as long as the interaction strength is not too large.

As an external perturbation, an X-ray field with frequency $\nu$ is coupled to the system with momentum-independent amplitude $W$,
\begin{equation}
H_X (t) = \frac{W}{\sqrt{V}} \sum_k e^{- \ii \nu t} a_k^\dagger a_d + \text{H.c.}
\end{equation}
$H_X (t)$ is chosen to describe only interband transitions because these are, in conjunction with the sharp Fermi surface and the flat lower band, responsible for the absorption singularity.
Correspondingly, we consider the X-ray frequency to be of the order of $|\varepsilon_d|$.
We use units with $\hbar = 1$.

A physical observable of interest is the X-ray absorption rate $R(\nu)$ or, equivalently, the excitation rate of the deep state.
When $\nu$ approaches the threshold frequency $\nu_c$, the leading behavior of $R(\nu)$ is a power-law divergence $\propto [\xi_0 / (\nu - \nu_c)]^{2 g}$ with $g = \rho U/V$.
This was conjectured by Mahan \cite{Mahan67} based on the terms up to third order of an expansion in powers of the interaction and later confirmed by Nozi\`eres and collaborators \cite{Roulet69,Nozieres69a}.
In linear response and for sufficiently small $|W|^2 / V$, the absorption rate can be accessed with many-body techniques via $R(\nu) = - 2|W|^2 \, \text{Im} \chi (\nu)$, where
\begin{equation}
\label{eq:ph_susceptibility}
\chi (\nu) = - \ii \frac{1}{V} \sum_{kk'} \int_{- \infty}^\infty \dd t \, e^{\ii \nu t} \left\langle \mathcal{T} a_d^\dagger (t) \, a_k (t) \, a_{k'}^\dagger (0) \, a_d (0) \right\rangle
\end{equation}
is a particle-hole susceptibility. Here, $\langle . \rangle$ denotes the ground-state expectation value, $a_{k/d}^{(\dagger)} (t)$ are the ladder operators in the Heisenberg picture with respect to the Hamiltonian (\ref{eq:Hamiltonian}), and $\mathcal{T}$ is the time-ordering operator.

A diagrammatic expansion of $\chi$ results in a power series in $U / V$.
Effectively, however, one obtains an expansion in powers of the dimensionless parameter $g = \rho U / V$ because for every additional interaction vertex in a diagram there is also one more independent momentum summation $\sum_k = \rho \int_{- \xi_0}^{\xi_0} \dd \varepsilon$.

We note that a many-body approach is not necessary to treat this model.
In fact, it has been solved exactly by applying a one-body scattering theory \cite{Nozieres69a}.
This is possible because the particular interaction term in Eq.~(\ref{eq:Hamiltonian}) does not alter the deep-state occupancy and acts just as a single-particle potential for the conduction states when the deep level is empty.
However, if one chooses to treat this model with many-body perturbation theory, one encounters the interesting problem of logarithmic divergences in two distinct channels (see also Sec.~\ref{sec:logarithmic_divergences}).
Being spinless and effectively zero-dimensional, it is probably the most basic model with this important feature.
Therefore, it was repeatedly used as a test bed to refine and compare various many-body approaches \cite{Roulet69,Nozieres69,Solyom74,Lange15,Kugler18a}.
Having an exact solution for comparison was then an additional advantage of this model.

If the system is prepared in a state with empty deep level, the X-ray field can induce the relaxation of an electron from the conduction band to the deep level.
This process is accompanied by X-ray emission.
In Ref.~\cite{Roulet69} the corresponding rate of stimulated X-ray emission is studied in close analogy to the rate of X-ray absorption within the zero-temperature formalism, see also our appendix \ref{sec:generalization_of_zero-temperature_formalism}.
On the leading-logarithmic level, the main part of the calculation turns out to be identical in both cases \cite{Roulet69}.
In this paper we focus on the case of absorption.
By following the arguments of Ref.~\cite{Roulet69}, all our considerations can be straightforwardly adapted to the case of emission.

\section{Perturbation theory within zero-temperature formalism}
\label{sec:perturbation_theory}

In this section we recap the most important features of a perturbative approach to the model.
Following largely Roulet et al.~\cite{Roulet69}, we choose the (real-time) zero-temperature formalism \cite{Negele88} as framework for the
diagrammatic expansion.
Our one-loop FRG approach developed in Sec.~\ref{sec:FRG_approach} is also formulated in the realm of this formalism.
This makes a detailed comparison between the parquet-based approach of Ref.~\cite{Roulet69} and the one-loop FRG approximation possible.

\subsection{Single-particle Green function}
\label{sec:single-particle_Green_function}

We choose to dress the propagator with the first-order contribution of the self-energy, resulting in
\begin{subequations}
\label{eq:propagators}
\begin{eqnarray}
\label{eq:deep-state_propagator} G_d^H (\omega) &=& \frac{1}{\omega - \tilde{\varepsilon}_d - \ii 0^+}\\
G_k^0 (\omega) &=& \frac{1}{\omega - \varepsilon_k + \ii 0^+ \sgn \varepsilon_k}
\end{eqnarray}
\end{subequations}
for the deep state and the conduction states, respectively.
Here, the deep-state Hartree self-energy has renormalized the deep level to $\tilde{\varepsilon}_d = \varepsilon_d + g \xi_0$.
No Fock contributions to the self-energy arise in this model: they would involve a free propagation between the deep state and a conduction state, which is not admitted by the free Hamiltonian.
For the conduction states, a single-particle perturbation $\propto - U / V$ arises when the interaction term in the Hamiltonian (\ref{eq:Hamiltonian}) is brought into the standard form by permuting all creation operators to the left.
It exactly cancels with the conduction-state Hartree self-energy.
This cancellation reflects that electrons in the conduction band do not interact with an electron occupying the deep state but only with a hole at the deep state; in fact, for this reason any (time-dependent) ground-state expectation value involving only conduction-state ladder operators is not affected by the interaction from Eq.~(\ref{eq:Hamiltonian}).
In the deep-state subspace, the Hartree-dressed propagator (\ref{eq:deep-state_propagator}) is analytic in the lower half-plane and thus purely advanced.
The same holds for the full deep-state propagator.
In time representation it takes the form
\begin{equation}
G_d (t) = - \ii \left\langle \mathcal{T} a_d (t) a_d^\dagger \right\rangle = \ii \Theta (-t) \left\langle a_d^\dagger a_d (t) \right\rangle
\end{equation}
so that it is directed backwards in time.
This can be understood as creating and subsequently annihilating a hole at the deep state.

In the following computations of two-particle quantities, we are not going to include additional self-energy contributions so that the propagator (\ref{eq:propagators}) will be used as the full single-particle Green function.
This is in fact correct for the conduction states because loops of two or more deep-state propagators vanish due to $G_d (t) \propto \Theta (-t)$, hence $G_k^0 = G_k$.
But in the case of the deep state, it is an approximation.
This does not influence the shape of the divergence of $\chi (\nu)$ as far as the leading logarithms are concerned \cite{Roulet69}.
However, it influences the threshold frequency which constitutes the position of the divergence and which is in our calculations $\nu_c = - \tilde{\varepsilon}_d = |\tilde{\varepsilon}_d|$.
The further discussion would also be possible after including other real, static contributions to the self-energy.
In that case only the specific value of the renormalized deep level $\tilde{\varepsilon}_d$ would differ.
Anyway, we are going to set $\tilde{\varepsilon}_d = 0$, see Sec.~\ref{sec:setting_renormalized_deep_level_to_zero}, and focus on investigating the shape of $\chi (\nu)$ near threshold.

The bare vertex has an incoming and outgoing leg for the deep state and an incoming and outgoing leg for the conduction states, but it has no actual dependence on the momenta.
Thus, all momentum summations are independent of each other and they can be performed immediately by employing the local conduction-electron propagator
\begin{subequations}
\begin{eqnarray}
\hspace{-1em} G_c (\omega) &=& \frac{1}{V} \sum_k G_k (\omega)\\
&=& \frac{\rho}{V} \left[ \ln \frac{|\xi_0 + \omega|}{|\xi_0 - \omega|} - \ii \pi \sgn (\omega) \Theta (\xi_0 - |\omega|) \right] .
\end{eqnarray}
\end{subequations}
In this sense the model is effectively zero-dimensional.
[In the exact analytic evaluation of diagrams, e.g., for Eq.~(\ref{eq:bare_bubbles}), it can still be helpful to integrate over frequencies before summing over momenta.]

\subsection{1PI two-particle vertex}

\begin{figure}
\includegraphics[scale=0.75]{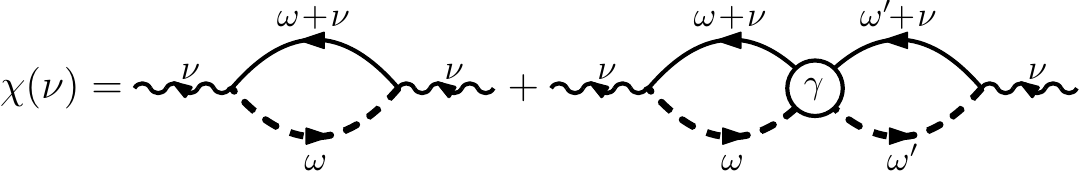}
\caption{\label{fig:ph-susceptibility}
Diagrammatic representation of Eq.~(\ref{eq:ph-susceptibility_with_1PI_vertex}).
Full lines refer to local conduction-electron propagators $G_c$, fat dashed lines to deep-state propagators $G_d$.
The circle stands for the 1PI two-particle vertex.
The three-leg vertices involving each a full, dashed, and wavy line conserve frequency, but do not contribute any factor.}
\end{figure}

The particle-hole susceptibility (\ref{eq:ph_susceptibility}), when expressed in terms of the 1PI two-particle vertex $\gamma$, can be calculated from
\begin{eqnarray}
\label{eq:ph-susceptibility_with_1PI_vertex}
\nonumber \chi (\nu) &=& - \ii \int \frac{\dd \omega}{2 \pi} \, G_d (\omega) G_c (\omega + \nu)\\
&& + \int \frac{\dd \omega \dd \omega'}{(2 \pi)^2} \, G_d (\omega) G_c (\omega + \nu) \bar{\gamma} (\omega, \omega'; \nu)\\
\nonumber && \hphantom{+ \int} \times G_d (\omega') G_c (\omega' + \nu).
\end{eqnarray}
The diagrammatic representation of this formula is shown in Fig.~\ref{fig:ph-susceptibility}.
Throughout this paper we draw full lines for local conduction-electron propagators and dashed lines for deep-state propagators.

\begin{figure}
\includegraphics[scale=0.75]{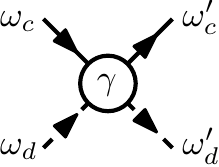}
\caption{\label{fig:1PI_vertex}
Diagrammatic representation of the 1PI vertex $\gamma_{d c|d c} (\omega_d', \omega_c'| \omega_d, \omega_c)$.
The external legs are meant to be amputated.
Frequency conservation assures $\omega_d + \omega_c = \omega_d' + \omega_c'$.
As independent frequencies we will employ $\omega_d, \omega_d'$, and either the total frequency $\Omega$ or the exchange frequency $X$, but not the conduction-state frequencies $\omega_c, \omega_c'$.}
\end{figure}

The 1PI vertex $\gamma_{d k'|d k}$ does not depend on the incoming and outgoing conduction-electron momentum $k$ and $k'$, respectively, because the interaction amplitude does not depend on the momenta, see Eq.~(\ref{eq:Hamiltonian}).
Therefore, we introduce the notation $\gamma_{d c|d c} = V \gamma_{d k'|d k}$.
For the frequency arguments, we employed in Eq.~(\ref{eq:ph-susceptibility_with_1PI_vertex}) the notation
\begin{subequations}
\label{eq:frequency_notation}
\begin{eqnarray}
\nonumber && \gamma_{d c|d c} (\omega_d', \omega_c'| \omega_d, \omega_c)\\
\label{eq:frequency_notation_pp} &=& 2 \pi \delta (\omega_d' + \omega_c' - \omega_d - \omega_c) \hat{\gamma} (\omega_d, \omega_d'; \Omega)\\
\label{eq:frequency_notation_ph} &=& 2 \pi \delta (\omega_d' + \omega_c' - \omega_d - \omega_c) \bar{\gamma} (\omega_d, \omega_d'; X),
\end{eqnarray}
\end{subequations}
see also Fig.~\ref{fig:1PI_vertex}.
Here, either the total frequency $\Omega = \omega_d + \omega_c = \omega'_d + \omega'_c$ or the exchange frequency $X = \omega'_c - \omega_d = \omega_c - \omega'_d$ has been chosen as one of the independent frequencies.
In Eq.~(\ref{eq:frequency_notation_ph}) we have chosen the order of the frequencies $\omega_d, \omega_d'$ in $\bar{\gamma} (\omega_d, \omega_d'; X)$ to match the order of the frequencies in Fig.~\ref{fig:ph-susceptibility}.

\subsection{Setting $\tilde{\varepsilon}_d = 0$}
\label{sec:setting_renormalized_deep_level_to_zero}

In the following calculation of $\chi (\nu)$, we set the renormalized deep level to $\tilde{\varepsilon}_d = 0$.
This is equivalent to measuring the X-ray frequency $\nu$ relative to the threshold frequency $|\tilde{\varepsilon}_d|$.
It is a convenient way to eliminate one of the parameters; the same was done by Roulet et al.~\cite{Roulet69}.
We present the reasoning behind this step by relating it to a Ward identity.
We intend to build on this brief discussion in a future publication, which addresses the same topic in the framework of the Matsubara formalism.

Let us consider a diagram contributing to $\chi (\nu) = \chi (\nu, \tilde{\varepsilon}_d)$ which arises when diagrams for the 1PI vertex and for the full deep-state lines are inserted into Fig.~\ref{fig:ph-susceptibility}.
We may choose the frequencies of the internal lines in accordance with frequency conservation such that the external frequency $\nu$ appears as addend in the frequency argument of every conduction-state propagator, but not of any deep-state propagator.
Subtracting a frequency $\alpha$ from the frequency arguments of all conduction- and deep-state propagators respects frequency conservation and does not alter the value of the diagram.
Then $\nu$ appears only in the conduction-state propagators, always in the form $\nu - \alpha$, and $\tilde{\varepsilon}_d$ appears only in the deep-state propagators, always in the form $\tilde{\varepsilon}_d + \alpha$.
This proves the Ward identity
\begin{equation}
\label{eq:zero-temperature_Ward_identity}
\chi (\nu, \tilde{\varepsilon}_d) = \chi (\nu - \alpha, \tilde{\varepsilon}_d + \alpha),
\end{equation}
which results from frequency conservation (i.e., time-translational invariance) and the conservation of the number of conduction- and deep-state electrons.

Equation (\ref{eq:zero-temperature_Ward_identity}) relates the susceptibilities of two models with respective values $\tilde{\varepsilon}_d$ and $\tilde{\varepsilon}_d + \alpha$ of the renormalized deep level.
These susceptibilities are defined in the state with a filled lower half of the conduction band and an occupied deep state.
For the model with deep-state energy $\tilde{\varepsilon}_d + \alpha$ and for sufficiently large $\alpha$, this is not the ground state of the interacting Hamiltonian.
Nonetheless, the zero-temperature formalism allows to compute expectation values in this state because it is an eigenstate of both the noninteracting and the interacting Hamiltonian, see appendix \ref{sec:generalization_of_zero-temperature_formalism}.
When $\chi$ is determined accordingly, the identity (\ref{eq:zero-temperature_Ward_identity}) holds for all real frequencies $\alpha$ and even on the level of individual diagrams.
As a special case, we obtain $\chi (\nu - \tilde{\varepsilon}_d, \tilde{\varepsilon}_d) = \chi (\nu , 0)$, where $\nu$ is now the deviation of the X-ray frequency from the threshold frequency $|\tilde{\varepsilon}_d|$.

\subsection{Logarithmic divergences}
\label{sec:logarithmic_divergences}

\begin{figure}
\includegraphics[scale=0.75]{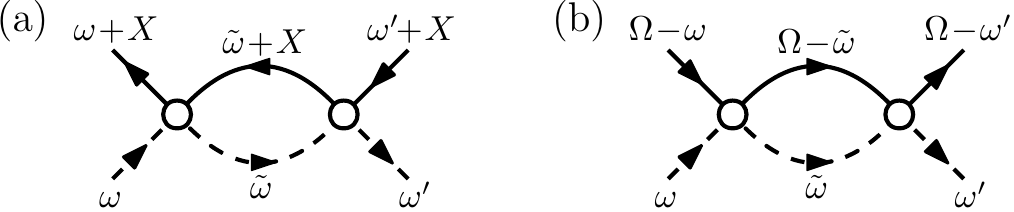}
\caption{\label{fig:bare_bubbles}
Second-order contributions to the 1PI vertex $\gamma_{d c| d c}$.
The small, empty circle represents the bare vertex contributing a factor $U$.
The thin dashed lines refer to Hartree-dressed deep-state propagators $G_d^H$.
In each channel the respective natural frequency has been employed.
(a) Particle-hole bubble depending on the exchange frequency $X$.
(b) Particle-particle bubble depending on the total frequency $\Omega$.}
\end{figure}

The second-order contributions to the 1PI vertex $\gamma_{d c| d c}$ are shown in Fig.~\ref{fig:bare_bubbles}.
They are the bare bubbles in the (exchange) particle-hole channel and in the particle-particle channel, which is strictly speaking a hole-hole channel, with the exact values
\begin{subequations}
\label{eq:bare_bubbles}
\begin{equation}
\label{eq:ph-bubble}
- g^2 \frac{V}{\rho} \left[ \ln \frac{|X|}{|\xi_0 - X|} - \ii \pi \, \Theta (\xi_0 - X) \Theta (X) \right]
\end{equation}
and
\begin{equation}
g^2 \frac{V}{\rho} \left[ \ln \frac{|\Omega|}{|\xi_0 + \Omega|} - \ii \pi \, \Theta (\xi_0 + \Omega) \Theta (- \Omega) \right] ,
\end{equation}
\end{subequations}
respectively.
As the frequencies at the external legs in Fig.~\ref{fig:bare_bubbles} have been chosen to already obey frequency conservation, there are no factors $2 \pi \delta (\ldots)$ in Eq.~(\ref{eq:bare_bubbles}).
The direct particle-hole bubble does not contribute to $\gamma_{d c| d c}$, but only to $\gamma_{d d| d d}$. The latter vertex is not considered here because its contribution is subleading \cite{Roulet69}.

Importantly, the bubbles (\ref{eq:bare_bubbles}) in both the (exchange) particle-hole and particle-particle channel feature a logarithmic divergence as their natural frequency $X$ or $\Omega$, respectively, approaches zero.
[There are also divergences for $X \to \xi_0$ and $\Omega \to - \xi_0$.
Those, however, turn out to be not important for $\chi (\nu)$ at small $\nu$.]
These diverging logarithms arise via the combination of the real part $\mathcal{P} \, 1 / \omega$ of the deep-state propagator with the discontinuous imaginary part $- \pi \rho \sgn (\omega) \Theta (\xi_0 - |\omega|) / V$ of the local conduction-electron propagator, e.g.,
\begin{eqnarray}
\label{eq:origin_of_logarithm}
\nonumber && \int \frac{\dd \tilde{\omega}}{2 \pi} \, \mathcal{P} \frac{1}{\tilde{\omega}} \, \sgn (\tilde{\omega} + X) \Theta (\xi_0 - |\tilde{\omega} + X|)\\
&=& - \frac{1}{\pi} \ln \frac{|X|}{\xi_0} + O \! \left( \frac{|X|}{\xi_0} \right)^2 .
\end{eqnarray}

\begin{figure}
\includegraphics[scale=0.75]{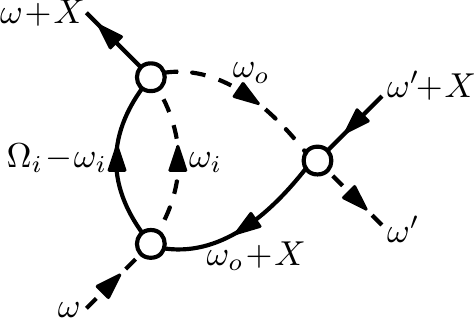}
\caption{\label{fig:crossed_bubble}
A third-order diagram contributing to $\bar{\gamma} (\omega, \omega'; X)$.
Dashed lines refer to Hartree-dressed deep-state propagators $G_d^H$.
From frequency conservation follows $\Omega_i = \omega_o + \omega + X$.}
\end{figure}

It is known that such logarithmic divergences appear for all powers of the interaction.
For the 1PI vertex, the \emph{leading} logarithms have the form $g^n \ln^{n-1}$ while subleading contributions contain at least one logarithm less.
Here, the arguments of the logarithms are essentially $|X| / \xi_0$ or $|\Omega| / \xi_0$ depending on the channel.
The arguments can also depend on the other free frequencies $\omega, \omega'$ if the corresponding external deep-state leg is not attached to the same vertex as one of the external conduction-state legs [see, e.g., Fig.~\ref{fig:crossed_bubble} and Eq.~(\ref{eq:crossed_bubble_result})].
All of the leading logarithms are contained within the parquet diagrams without any self-energy insertions (except for static contributions absorbed into $\tilde{\varepsilon}_d$) \cite{Roulet69}.
These diagrams can be constructed by starting with the bare vertex and successively replacing any vertex with either of the bubbles given in Fig.~\ref{fig:bare_bubbles}.

The leading logarithms appearing in the expansion of the particle-hole susceptibility $\chi (\nu)$, which directly follow from those of $\gamma$ via Eq.~(\ref{eq:ph-susceptibility_with_1PI_vertex}), assume the form $g^n \ln^{n+1} (|\nu| / \xi_0)$.
In comparison to the 1PI vertex, two additional powers of the logarithm arise from the two external bubbles in the right diagram in Fig.~\ref{fig:ph-susceptibility}.
Close to the threshold $\nu = 0$ where $g \ln (|\nu| / \xi_0)$ is not much smaller than one, these terms are significant for arbitrary powers $n$ even though $g$ itself is small.
To approximate the behavior of $\chi (\nu)$ in a reasonable way, it is then necessary to resum the leading logarithms of all orders.
Even closer to the threshold, i.e., as $\nu$ goes to zero, $g \ln (|\nu| / \xi_0)$ increases further until also subleading logarithms become large and must be included.
However, in this paper only the resummation of the leading logarithms is discussed.

As an example consider the zeroth-order contribution to $\chi (\nu)$, which is contained in the first addend on the right-hand side of Eq.~(\ref{eq:ph-susceptibility_with_1PI_vertex}).
It can be obtained from the result of the particle-hole bubble given in Eq.~(\ref{eq:ph-bubble}) with $X = \nu$ by replacing the prefactor of the square brackets with $\rho / V$.
The leading logarithm $g^0 \ln(|\nu| / \xi_0)$ then appears only in the real part.
In fact, the leading logarithms of all orders appear only in the real part of the particle-hole susceptibility \cite{Roulet69}.
If one employs a scheme to capture just the leading logarithms, one therefore has to recover the imaginary part in order to determine the absorption rate.
This can be done as outlined by Roulet et al.~\cite{Roulet69}.

It is important to note that it is not necessary to include the exact value of any given parquet diagram in the resummation scheme in order to obtain a valid leading-logarithmic approximation.
Instead, one can make further approximations as long as they do not influence the leading logarithms.
That the parquet diagrams indeed contain subleading contributions is already obvious from the exact results (\ref{eq:bare_bubbles}) for the bubbles, where, e.g.,
\begin{equation}
\ln \frac{|X|}{|\xi_0 - X|} = \ln \frac{|X|}{\xi_0} - \ln \frac{|\xi_0 - X|}{\xi_0}.
\end{equation}

\subsection{Third-order contribution as key example}
\label{sec:third-order_example}

Roulet et al. worked out how to extract the leading contribution from a given parquet diagram \cite{Roulet69}.
We briefly recap their scheme by applying it to the third-order diagram of $\bar{\gamma} (\omega, \omega'; X)$ shown in Fig.~\ref{fig:crossed_bubble}.
The achieved insights will form the basis for the construction of an RG treatment in Sec.~\ref{sec:FRG_approach}.

The parts of the propagators that do not give rise to the diverging logarithm in a bare bubble are neglected, i.e., only the real part of the deep-state propagator and the imaginary part of the local conduction-electron propagator are retained.
The diagram in Fig.~\ref{fig:crossed_bubble} then translates into
\begin{eqnarray}
\label{eq:crossed_bubble_expression}
\nonumber c \int \frac{\dd \omega_o \, \dd \omega_i}{(2 \pi)^2} \mathcal{P} \frac{1}{\omega_o} \sgn (\omega_o + X) \Theta (\xi_0 - |\omega_o + X|)\\
\times \mathcal{P} \frac{1}{\omega_i} \sgn (\Omega_i - \omega_i) \Theta(\xi_0 - |\Omega_i - \omega_i|)
\end{eqnarray}
with the prefactor $c = \pi^2 g^3 V / \rho$ and with the abbreviation $\Omega_i = \Omega_i (\omega_o, \omega, X) = \omega_o + \omega + X$.

The indices $i$ and $o$ refer to the ``inner'' and ``outer'' bubble of the diagram, respectively.
When (a part of) a diagram can be constructed by replacing a vertex at an end of some bubble by another bubble (or chain of bubbles) of the opposite channel, then we call the latter bubble (or chain of bubbles) the inner one and the former bubble the outer one.
Repeating this construction establishes the strict partial order ``being inner to'' among the bubbles of a parquet diagram.
At the end of this subsection, we will conclude that there is a related order among the absolute values of the deep-state frequencies of the bubbles as far as the leading-logarithmic approximation is concerned.

The integral over the frequency of the inner bubble in Eq.~(\ref{eq:crossed_bubble_expression}) yields
\begin{subequations}
\begin{eqnarray}
\nonumber && \int \frac{\dd \omega_i}{2 \pi} \, \mathcal{P} \frac{1}{\omega_i} \sgn(\Omega_i - \omega_i) \Theta(\xi_0 - |\Omega_i - \omega_i|)\\
\label{eq:internal_bubble_short_result} &=& \frac{1}{\pi} \ln \frac{|\Omega_i|}{\sqrt{|\xi_0^2 - \Omega_i^2|}}\\
\label{eq:internal_bubble_break_down} &=& \frac{1}{\pi} \left( \ln \frac{M}{\xi_0} + \ln \frac{|\Omega_i|}{M} + \ln \frac{\xi_0}{\sqrt{|\xi_0^2 - \Omega_i^2|}} \right)
\end{eqnarray}
\end{subequations}
with $M = M(\omega_o, \omega, X) = \max \{ |\omega_o|, |\omega|, |X| \}$.
In the particular case $|X| < |\omega| \ll \xi_0$, this result can be approximated by $[\ln (M / \xi_0)] / \pi$ when inserted into Eq.~(\ref{eq:crossed_bubble_expression}): the other two logarithmic addends contribute only subleadingly, cf. Ref.~\cite{Roulet69}.
The leading-logarithmic approximation to the value of the diagram is hence
\begin{eqnarray}
\label{eq:crossed_bubble_result}
\nonumber && \frac{c}{2 \pi^2} \int_{- \xi_0 - X}^{\xi_0 - X} \dd \omega_o \mathcal{P} \frac{1}{\omega_o} \sgn (\omega_o + X) \ln \frac{\max \{ |\omega_o|, |\omega| \} }{\xi_0}\\
& \approx & - \frac{1}{2} g^3 \frac{V}{\rho} \ln \frac{|\omega|}{\xi_0} \left( \ln \frac{|X|}{|\omega|} + \ln \frac{|X|}{\xi_0} \right)
\end{eqnarray}
for $|X| < |\omega| \ll \xi_0$, a result which is $\propto g^3 \ln^2$ as expected.

It is illuminating to identify the particular subregion of frequency integration that is responsible for this leading-logarithmic result.
For $|\Omega_i| \le \xi_0 / 2$ the range $\omega_i \in [- |\Omega_i|, |\Omega_i|]$ does not contribute to the value of the integral in Eq.~(\ref{eq:internal_bubble_short_result}); this results from the combination of the principal value and the sign function in the integrand.
Similarly, approximating the integral by $[\ln (M / \xi_0)] / \pi$ means to restrict the range of integration to $M < |\omega_i| < \xi_0$.
The leading contribution actually results from the small frequencies with $M < |\omega_i| \ll \xi_0$.
Larger $|\omega_i|$ are not important, e.g., the range $\xi_0 / 10 < |\omega_i| < \xi_0$ yields only the subleading contribution $- (\ln 10) / \pi$.
Very similarly, the relevant integration range of the frequency $\omega_o$ is $|X| < |\omega_o| \ll \xi_0$.
Two important conclusions can be drawn from these observations.

Firstly, it is indeed sufficient to consider only the case $|X| < |\omega| \ll \xi_0$.
If the diagram in Fig.~\ref{fig:crossed_bubble} is inserted for the 1PI vertex in the representation of $\chi (\nu)$ shown in Fig.~\ref{fig:ph-susceptibility}, $X$ assumes the value of the X-ray frequency $\nu$; the leading-logarithmic behavior near threshold, which we are interested in, emerges then for $|X| = |\nu| \ll \xi_0$.
Furthermore, when the argument used above for $\omega_i$ and $\omega_o$ is applied to the additional $\omega$-integration that appears in the diagram for $\chi (\nu)$, it shows that the relevant frequencies $\omega$ are from the range $|X| < |\omega| \ll \xi_0$ as well.
The same reasoning is possible if the diagram in Fig.~\ref{fig:crossed_bubble} is not directly inserted for the 1PI vertex in Fig.~\ref{fig:ph-susceptibility}, but is used as part of a larger parquet diagram which in turn is inserted for that 1PI vertex.

Secondly, the restriction to $M < |\omega_i|$ with $M = \max \{ |\omega_o|, |\omega| \}$ implies $|\omega_o| < |\omega_i|$.
For the leading logarithmic contribution, it hence suffices to integrate with respect to the deep-state frequency of the inner bubble over only those regions where its absolute value is greater than the one of the deep-state frequency of the outer bubble.
This statement can be generalized to all parquet diagrams and all pairs of bubbles where one is inner to the other; corresponding observations are described in Ref.~\cite{Abrikosov65,Binz03}.
This is the very property of the parquet diagrams that is responsible for the success of our \emph{one-loop} FRG approach to reproduce the leading-logarithmic approximation.

\section{FRG in zero-temperature formalism for a general model}
\label{sec:FRG_general_models}

In this section we develop a formulation of the FRG method \cite{Metzner12, Kopietz10, Platt13} in the framework of the (real-time) zero-temperature formalism, also known as ground-state formalism.
Since we are not aware of an FRG scheme in the literature that is based on this formalism, we present the derivation of FRG flow equations for a general model of interacting fermions.
Readers who are not interested in details on how to establish a zero-temperature FRG can skip this section.
Its central result that is subsequently used in Sec.~\ref{sec:FRG_approach} are the flow equations given in Eq.~(\ref{eq:general_flow_of_self-energy}) and (\ref{eq:general_flow_of_two-particle_vertex}).

The FRG flow equations for a class of correlation functions, e.g., Green functions or 1PI vertex functions, can be derived from the corresponding generating functional.
We will start by deriving a functional-integral representation of a generating functional of Green functions for an interacting system in the ground state.

In Ref.~\cite{Negele88} such a functional-integral representation is presented, but only for the noninteracting case.
There, the derivation is based on a non-standard variant of coherent states -- namely the common eigenstates of the annihilators of single-particle states that are empty in the noninteracting ground state and of the creators of single-particle states that are occupied in the noninteracting ground state.
Compared to standard coherent states, which are the common eigenstates of all annihilators, the role of creators and annihilators has been swapped for levels below the Fermi energy.
As a consequence the noninteracting ground state acquires the role of the vacuum state.
While this approach allows for an elegant functional-integral representation of the generating functional in the noninteracting case, see Ref.~\cite{Negele88}, we found it rather tedious to work with the corresponding representation in the interacting case: treating the coherent-state matrix elements of the interaction turns out to be cumbersome.
We consider this to be a drawback not only regarding the discussion of the FRG flow equations but also regarding the derivation of the diagrammatic perturbation theory within the functional-integral formulation.

In contrast to Ref.~\cite{Negele88}, we use standard coherent states, which turns out to be straightforward.
However, we follow Ref.~\cite{Negele88} in regard to deriving the ground-state expectation value from a damped time evolution instead of using the Gell-Mann and Low theorem.

\subsection{Definition of Green functions and their generating functional}
\label{sec:definition_of_Green_functions_and_their_generating_functional}

We consider a general Hamiltonian for an interacting many-fermion system,
\begin{subequations}
\label{eq:general_Hamiltonian}
\begin{align}
H &= H_0 + H_\text{int}\\
&= \sum_\alpha \varepsilon_\alpha a_\alpha^\dagger a_\alpha + \frac{1}{4} \sum_{\alpha'_1 \alpha'_2 \alpha_1 \alpha_2} \bar{v}_{\alpha'_1 \alpha'_2 \alpha_1 \alpha_2} a_{\alpha'_1}^\dagger a_{\alpha'_2}^\dagger a_{\alpha_2} a_{\alpha_1},
\end{align}
\end{subequations}
where $\alpha = 1, 2, \ldots$ numbers the single-particle eigenstates of $H_0$ such that the eigenenergies are ordered monotonically $\varepsilon_1 \le \varepsilon_2 \le \ldots$
Let the particle number $N$ be fixed and let there be a gap $\varepsilon_N < \varepsilon_{N + 1}$.
Then the ground state of the noninteracting Hamiltonian $H_0$ is nondegenerate and given by $| \Phi_0 \rangle = a_1^\dagger \ldots a_N^\dagger | 0 \rangle$ with $| 0 \rangle$ being the vacuum state.
We choose the zero of single-particle energies to lie between $\varepsilon_N$ and $\varepsilon_{N + 1}$ so that the negative levels $\varepsilon_1, \ldots, \varepsilon_N < 0$ are occupied and the positive levels $\varepsilon_{N + 1}, \dots > 0$ are empty in the noninteracting ground state.
The corresponding occupation numbers are
\begin{equation}
n_\alpha = \left\langle \Phi_0 \middle| a_\alpha^\dagger a_\alpha \middle| \Phi_0 \right\rangle = \begin{cases} 1, & \alpha \le N\\ 0, & \alpha > N. \end{cases}
\end{equation}
The normalized ground state of the interacting Hamiltonian $H$ shall be denoted by $| \Psi_0 \rangle$.
It is assumed to be nondegenerate and not orthogonal to $| \Phi_0 \rangle$.

Note that this scenario applies also to the model of X-ray absorption in metals even though that model involves a continuous conduction band.
An integration over said band is just an approximation for the summation over a rather dense but discrete spectrum of plane-wave states.
In particular, the noninteracting and interacting ground states are nondegenerate and not mutually orthogonal; in fact, they are identical, see also appendix \ref{sec:generalization_of_zero-temperature_formalism}.

The time-ordered multi-particle Green functions are defined as
\begin{eqnarray}
\label{eq:definition_of_Green_functions}
&& G(\alpha_1 t_1, \ldots, \alpha_n t_n| \alpha'_1 t'_1, \ldots, \alpha'_n t'_n)\\
\nonumber &=& (- i)^n \! \left\langle \Psi_0 \middle| \mathcal{T} a_{\alpha_1} \! (t_1) \ldots a_{\alpha_n} \! (t_n) a_{\alpha'_n}^\dagger \! (t'_n) \ldots a_{\alpha'_1}^\dagger \! (t'_1) \middle| \Psi_0 \right\rangle \! .
\end{eqnarray}
Similarly to the discussion in Ref.~\cite{Negele88}, one finds that they can be determined from a damped time evolution.
This can formally be realized via $G = \lim_{\eta \to 0^+} G_\eta$ and
\begin{eqnarray}
\label{eq:Green_functions_from_generating_functional}
&& G_\eta (\alpha_1 t_1, \ldots, \alpha_n t_n| \alpha'_1 t'_1, \ldots, \alpha'_n t'_n)\\
\nonumber &=& (- i)^n \frac{\delta^{2n} \mathcal{G}_\eta [\bar{J}, J]}{\delta \bar{J}_{\alpha_1} \! (t_1) \ldots \delta \bar{J}_{\alpha_n} \! (t_n) \delta J_{\alpha'_n} \! (t'_n) \ldots \delta J_{\alpha'_1} \! (t'_1)} \Bigg|_{\bar{J} = 0 = J}
\end{eqnarray}
with the generating functional
\begin{equation}
\label{eq:generating_functional_definition}
\mathcal{G}_\eta [\bar{J}, J] = \lim_{t_0 \to \infty} \frac{Z_\eta [\bar{J}, J]}{Z_\eta [0, 0]}
\end{equation}
and
\begin{equation}
\label{eq:Z_definition}
Z_\eta [\bar{J}, J] = \big\langle \Phi_0 \big| U^{(\eta)}_{\bar{J}, J} (t_0, - t_0) \big| \Phi_0 \big\rangle.
\end{equation}
In the time evolution operator
\begin{eqnarray}
\label{eq:time_evolution_operator}
&& U^{(\eta)}_{\bar{J}, J} (t_0, - t_0)\\
\nonumber & \! =& \mathcal{T} \! \exp \! \Bigg( \! \! \! - \! \ii \! \int_{- t_0}^{t_0} \! \! \dd t \bigg\{ \! (1 \! - \! \ii \eta) H \! + \! \sum_\alpha \! \big[ \bar{J}_\alpha (t) a_\alpha \! + \! a_\alpha^\dagger J_\alpha (t) \big] \! \bigg\} \! \Bigg),
\end{eqnarray}
source terms with external Grassmann variables $\bar{J}_\alpha$, $J_\alpha$ were added to the Hamiltonian.
When $| \Phi_0 \rangle$ is expanded in eigenstates of the interacting Hamiltonian $H$, the factor $1 - \ii \eta$ with $\eta > 0$ suppresses the contributions from excited states to the Green functions, leaving only the ground-state expectation value as required by the definition (\ref{eq:definition_of_Green_functions}).
In contrast to Ref.~\cite{Negele88}, in which a time contour in the complex plane is used, we formally attribute this factor not to the time integration, but to the Hamiltonian.
This corresponds to the picture that excited states decay due to a non-zero anti-Hermitian part of the Hamiltonian.

\subsection{Discrete integral expression for $Z_\eta [\bar{J}, J]$}

We introduce intermediate time steps $\tau_m = - t_0 + m \Delta$ with $\Delta = 2 t_0 / M$ and $m = 0, 1, \ldots, M$ and insert resolutions of unity into Eq.~(\ref{eq:Z_definition}) in terms of standard fermionic coherent states
\begin{equation}
| \varphi \rangle = \exp \left( - \sum_\alpha \varphi_\alpha a_\alpha^\dagger \right) | 0 \rangle,
\end{equation}
where $\varphi$ stands for the set of Grassmann generators $\{ \varphi_1, \varphi_2, \ldots \}$.
This yields
\begin{eqnarray}
\label{eq:Z_with_resolutions_of_unity}
\nonumber Z_\eta [\bar{J}, J] &=& \int \left( \prod_{m = 0}^M \prod_\alpha \dd \bar{\varphi}^m_\alpha \dd \varphi^m_\alpha e^{- \bar{\varphi}^m_\alpha \varphi^m_\alpha} \right) \big\langle \Phi_0 \big| \varphi^M \big\rangle\\
&& \hphantom{\int} \times \big\langle \varphi^M \big| U^{(\eta)}_{\bar{J}, J} (\tau_M, \tau_{M - 1}) \big| \varphi^{M - 1} \big\rangle\\
\nonumber && \hphantom{\int} \times \ldots \, \big\langle \varphi^1 \big| U^{(\eta)}_{\bar{J}, J} (\tau_1, \tau_0) \big| \varphi^0 \big\rangle \; \big\langle \varphi^0 \big| \Phi_0 \big\rangle,
\end{eqnarray}
where each $\bar{\varphi}^m_\alpha$ is an additional Grassmann generator that is by definition the conjugate of $\varphi^m_\alpha$.
The usage of standard coherent states is an important difference to Ref.~\cite{Negele88}.
It will allow for a straightforward derivation of the functional-integral representation of the generating functional in the interacting case, see Eq.~(\ref{eq:generating_functional_continuous}) below.
The factors $\big\langle \Phi_0 \big| \varphi^M \big\rangle = \varphi^M_N \ldots \varphi^M_1$ and $\big\langle \varphi^0 \big| \Phi_0 \big\rangle = \bar{\varphi}^0_1 \ldots \bar{\varphi}^0_N$ in Eq.~(\ref{eq:Z_with_resolutions_of_unity}) are important for the form of the free propagator, see the integration in Eq.~(\ref{eq:reduction_of_Grassmann_integrals}) and the remark at the end of Sec.~\ref{sec:noninteracting_generating_functional}.
Up to corrections $\propto \Delta^2$, the occurring matrix elements are given by
\begin{subequations}
\label{eq:matrix_elements}
\begin{eqnarray}
\nonumber && \left\langle \varphi^m \middle| U^{(\eta)}_{\bar{J}, J} (\tau_m, \tau_{m - 1}) \middle| \varphi^{m - 1} \right\rangle\\
\nonumber &=& \exp \bigg( \sum_\alpha \bar{\varphi}^m_\alpha \varphi^{m - 1}_\alpha \bigg)\\
\label{eq:intermediate_matrix_elements} && \times \bigg\{ 1 - \ii \Delta \bigg[ (1 - \ii \eta) H(\bar{\varphi}^m, \varphi^{m - 1})\\
\nonumber && \hphantom{\times \bigg\{ 1 - \ii \Delta \bigg[} + \sum_\alpha \left( \bar{J}^{m - 1}_\alpha \varphi^{m - 1}_\alpha + \bar{\varphi}^m_\alpha J^m_\alpha \right) \bigg] \bigg\}\\
\nonumber &=& \exp \bigg\{ \sum_\alpha \bar{\varphi}^m_\alpha e^{- (\ii + \eta) \varepsilon_\alpha \Delta} \varphi^{m - 1}_\alpha\\
&& \hphantom{\exp \bigg\{} - \ii \Delta \bigg[ (1 - \ii \eta) H_\text{int} (\bar{\varphi}^m, \varphi^{m - 1})\\
\nonumber && \hphantom{\exp \bigg\{ - \ii \Delta \bigg[} + \sum_\alpha \left( \bar{J}^{m - 1}_\alpha \varphi^{m - 1}_\alpha + \bar{\varphi}^m_\alpha J^m_\alpha \right) \bigg] \bigg\},
\end{eqnarray}
\end{subequations}
where we used the notation $\bar{J}^m_\alpha = \bar{J}_\alpha (\tau_m)$ and $J^m_\alpha = J_\alpha (\tau_m)$.
The expression for $H(\bar{\varphi}^m, \varphi^{m - 1})$ can be obtained from Eq.~(\ref{eq:general_Hamiltonian}) by replacing all ladder operators with Grassmann generators according to $a_\alpha^\dagger \to \bar{\varphi}^m_\alpha$ and $a_\alpha \to \varphi^{m - 1}_\alpha$.
In particular, we have
\begin{eqnarray}
\nonumber && H_\text{int} (\bar{\varphi}^m, \varphi^{m - 1})\\
&=& \frac{1}{4} \sum_{\alpha'_1 \alpha'_2 \alpha_1 \alpha_2} \bar{v}_{\alpha'_1 \alpha'_2 \alpha_1 \alpha_2} \bar{\varphi}^m_{\alpha'_1} \bar{\varphi}^m_{\alpha'_2} \varphi^{m - 1}_{\alpha_2} \varphi^{m - 1}_{\alpha_1}.
\end{eqnarray}
(If one uses the particular coherent states of Ref.~\cite{Negele88} instead, the expression that results for $H_\text{int}$ is not as simple.)
Since none of the matrix elements (\ref{eq:matrix_elements}) depend on $\bar{\varphi}^0$ or $\varphi^M$, the integrations for $m = 0, M$ in Eq.~(\ref{eq:Z_with_resolutions_of_unity}) reduce to
\begin{subequations}
\label{eq:reduction_of_Grassmann_integrals}
\begin{eqnarray}
\nonumber && \int \! \Bigg( \! \prod_\alpha \dd \bar{\varphi}^0_\alpha \dd \varphi^0_\alpha e^{- \bar{\varphi}^0_\alpha \varphi^0_\alpha} \dd \bar{\varphi}^M_\alpha \dd \varphi^M_\alpha e^{- \bar{\varphi}^M_\alpha \varphi^M_\alpha} \! \Bigg)\\
\nonumber && \hphantom{\int \!} \times \varphi^M_N \! \ldots \varphi^M_1 \bar{\varphi}^0_1 \ldots \bar{\varphi}^0_N f \big( \bar{\varphi}^M, \varphi^0 \big)\\
\nonumber & \! \! =& \int \! \Bigg( \! \prod_{\alpha \le N} \! \dd \bar{\varphi}^M_\alpha \dd \varphi^M_\alpha \! \! \Bigg) \varphi^M_N \! \ldots \varphi^M_1 \Bigg( \! \prod_{\alpha \le N} \! \dd \bar{\varphi}^0_\alpha \dd \varphi^0_\alpha \! \Bigg) \bar{\varphi}^0_1 \ldots \bar{\varphi}^0_N\\
\nonumber && \hphantom{\int \!} \times \! \Bigg[ \prod_{\alpha > N} \! \dd \bar{\varphi}^0_\alpha \dd \varphi^0_\alpha \big( 1 - \bar{\varphi}^0_\alpha \varphi^0_\alpha \big) \dd \bar{\varphi}^M_\alpha \dd \varphi^M_\alpha \big( 1 - \bar{\varphi}^M_\alpha \varphi^M_\alpha \big) \Bigg]\\
&& \hphantom{\int \!} \times f \big( \bar{\varphi}^M, \varphi^0 \big)\\
& \! \! =& (- 1)^N \int \! \Bigg( \! \prod_{\alpha \le N} \! \dd \bar{\varphi}^M_\alpha \dd \varphi^0_\alpha \! \Bigg) f \big( \bar{\varphi}^M, \varphi^0 \big) \bigg|_{\mathcal{B}_>}.
\end{eqnarray}
\end{subequations}
The notation in the last line involving the boundary conditions
\begin{equation}
\mathcal{B}_> = \left\{ \bar{\varphi}^M_{\alpha > N} = 0, \varphi^0_{\alpha > N} = 0 \right\}
\end{equation}
means that in $f \big( \bar{\varphi}^M, \varphi^0 \big)$ the generators $\bar{\varphi}^M_\alpha$, $\varphi^0_\alpha$ with $\alpha > N$ are replaced by zero.
These boundary conditions reflect that the levels with $\alpha > N$ are empty in the state $| \Phi_0 \rangle$.
In total one obtains
\begin{equation}
\label{eq:Z_discrete_Grassmann_integral}
Z_\eta [\bar{J}, J] = \lim_{M \to \infty} \int D_M (\bar{\varphi}, \varphi) e^{\ii S_M (\bar{\varphi}, \varphi; \bar{J}, J)}
\end{equation}
with the Grassmann integration measure
\begin{eqnarray}
&& D_M (\bar{\varphi}, \varphi)\\
\nonumber &=& (-1)^{M N} \Bigg( \! \prod_{\alpha \le N} \prod_{m = 1}^M \! \dd \bar{\varphi}^m_\alpha \dd \varphi^{m - 1}_\alpha \! \Bigg) \Bigg( \! \prod_{\alpha > N} \prod_{m = 1}^{M - 1} \! \dd \bar{\varphi}^m_\alpha \dd \varphi^m_\alpha \! \Bigg).
\end{eqnarray}
The action $S_M (\bar{\varphi}, \varphi; \bar{J}, J)$ is the sum of the free part
\begin{eqnarray}
\label{eq:free_part_of_action_discrete}
\nonumber S_M^0 (\bar{\varphi}, \varphi) &=& \sum_{\alpha', \alpha \le N} \sum_{m' = 1}^M \sum_{m = 0}^{M - 1} \bar{\varphi}^{m'}_{\alpha'} Q^{m' m}_{\alpha' \alpha} \varphi^{m}_\alpha\\
&& + \sum_{\alpha', \alpha > N} \sum_{m', m = 1}^{M - 1} \bar{\varphi}^{m'}_{\alpha'} Q^{m' m}_{\alpha' \alpha} \varphi^m_\alpha
\end{eqnarray}
with
\begin{equation}
\label{eq:inverse_free_propagator}
Q^{m' m}_{\alpha' \alpha} = \ii \delta_{\alpha' \alpha} \left[ \delta_{m' m} - \delta_{m' - 1, m} e^{- (\ii + \eta) \varepsilon_\alpha \Delta} \right],
\end{equation}
the interaction part
\begin{equation}
S_M^\text{int} (\bar{\varphi}, \varphi) = - (1 - \ii \eta) \Delta \sum_{m = 1}^M H_\text{int} (\bar{\varphi}^m, \varphi^{m - 1}) \bigg|_{\mathcal{B}_>},
\end{equation}
and the source part
\begin{eqnarray}
\label{eq:source_part_of_action_discrete}
\nonumber S_M^\text{source} (\bar{\varphi}, \varphi; \bar{J}, J) &=& - \Delta \! \sum_{\alpha \le N} \sum_{m = 1}^M \! \big( \bar{J}^{m - 1}_\alpha \varphi^{m - 1}_\alpha + \bar{\varphi}^m_\alpha J^m_\alpha \big)\\
&& - \Delta \! \sum_{\alpha > N} \sum_{m = 1}^{M - 1} \! \big( \bar{J}^m_\alpha \varphi^m_\alpha + \bar{\varphi}^m_\alpha J^m_\alpha \big) .
\end{eqnarray}

\subsection{Noninteracting generating functional}
\label{sec:noninteracting_generating_functional}

In the noninteracting case, the integral in Eq.~(\ref{eq:Z_discrete_Grassmann_integral}) is of Gaussian form.
We consider $Q$ given by Eq.~(\ref{eq:inverse_free_propagator}) to be a matrix and introduce row vectors $\bar{\varphi}$, $\bar{J}$ and column vectors $\varphi$, $J$.
We point out the peculiar ranges of the discrete-time indices [see Eq.~(\ref{eq:free_part_of_action_discrete}) and (\ref{eq:source_part_of_action_discrete})]:
In the sector with $\alpha \le N$, the row index $m'$ of $Q$ runs from $1$ to $M$, whereas its column index $m$ runs from $0$ to $M - 1$.
Correspondingly, the indices of $\bar{\varphi}$ and $J$ run from $1$ to $M$, whereas those of $\varphi$ and $\bar{J}$ run from $0$ to $M - 1$.
In the sector with $\alpha > N$, all discrete-time indices simply run from $1$ to $M - 1$.
The Gaussian integral evaluates to
\begin{eqnarray}
\label{eq:Gaussian_integral}
\nonumber && \int D_M (\bar{\varphi}, \varphi) e^{\ii [\bar{\varphi} Q \varphi - \Delta (\bar{J} \varphi + \bar{\varphi} J)]}\\
&=& e^{- (\ii + \eta) 2 t_0 \sum_{\alpha \le N} \varepsilon_\alpha} \; e^{- \ii \Delta^2 \bar{J} Q^{-1} J}.
\end{eqnarray}
The result for the noninteracting generating functional is thus
\begin{equation}
\mathcal{G}_\eta^0 [\bar{J}, J] = \lim_{t_0 \to \infty} \lim_{M \to \infty} e^{- \ii \Delta^2 \bar{J} g J}
\end{equation}
with the free propagator
\begin{subequations}
\begin{eqnarray}
g^{m m'}_{\alpha \alpha'} &=& (Q^{-1})^{m m'}_{\alpha \alpha'}\\
\nonumber &=& - \ii \delta_{\alpha \alpha'} e^{- (\ii + \eta) \varepsilon_\alpha (\tau_m - \tau_{m'})}\\
&& \times \begin{cases} 1 - n_\alpha, & M - 1 \ge m \ge m' \ge 1\\ - n_\alpha, & 0 \le m < m' \le M. \end{cases}
\end{eqnarray}
\end{subequations}

The free propagator is purely advanced for $\alpha \le N$ and purely retarded for $\alpha > N$.
The inverse of $Q$ assumes such distinct forms in the two sectors because of the differently restricted ranges of the discrete-time indices.
These in turn are a consequence of the integrations over the Grassmann generators at the boundaries, which were performed in Eq.~(\ref{eq:reduction_of_Grassmann_integrals}).

\subsection{Continuous notation}
\label{sec:continuous_notation}

In the limit $M \to \infty$, the free propagator becomes
\begin{subequations}
\label{eq:free_propagator_continuous}
\begin{eqnarray}
g_{\alpha \alpha'} (t, t') &=& \delta_{\alpha \alpha'} g_\alpha (t - t')\\
\nonumber g_\alpha (t) &=& - \ii e^{- (\ii + \eta) \varepsilon_\alpha t} \big[ (1 - n_\alpha) \Theta (t - 0^+)\\
&& \hphantom{- \ii e^{- (\ii + \eta) \varepsilon_\alpha t} \big[} - n_\alpha \Theta (- t + 0^+) \big] .
\end{eqnarray}
\end{subequations}
Following the usual convention \cite{Negele88}, we have chosen $g_\alpha (0) = g_\alpha (0^-)$.
This choice is advantageous for the diagrammatic expansion: It will allow to drop the infinitesimal differences of the times at each vertex, see Eq.~(\ref{eq:bare_vertex_general}), which matter only if a free propagator connects a vertex with itself.
And if two external ladder operators in Eq.~(\ref{eq:definition_of_Green_functions}) happen to be at equal times and to be paired by Wick's theorem, the choice agrees with the property $\mathcal{T} a_\alpha (t) a_{\alpha'}^\dagger (t) = - a_{\alpha'}^\dagger (t) a_\alpha (t)$ of the time-ordering operator.

Also on the level of the action and of the integral expression (\ref{eq:Z_discrete_Grassmann_integral}), it is possible to go over to a continuous notation.
The details are shown in appendix \ref{sec:details_on_continuum_limit}.
For the generating functional, we obtain the functional-integral representation
\begin{eqnarray}
\label{eq:generating_functional_continuous}
&& \mathcal{G}_\eta [\bar{J}, J]\\
\nonumber &=& \frac{\int D[\bar{\varphi}, \varphi] \exp \big\{ \ii \bar{\varphi} Q \varphi + \ii S^\text{int} [\bar{\varphi}, \varphi] - \ii (\bar{J} \varphi + \bar{\varphi} J) \big\} }{\int D[\bar{\varphi}, \varphi] \exp \big\{ \ii \bar{\varphi} Q \varphi + \ii S^\text{int} [\bar{\varphi}, \varphi] \big\} },
\end{eqnarray}
where $Q$ is now the differential operator given by Eq.~(\ref{eq:inverse_free_propagator_continuous}) and where the interaction part of the action can be written as
\begin{equation}
S^\text{int} [\bar{\varphi}, \varphi] = - \frac{1}{4} \sum_{x'_1 x'_2 x_1 x_2} \bar{v}_{x'_1 x'_2 x_1 x_2} \bar{\varphi}_{x'_1} \bar{\varphi}_{x'_2} \varphi_{x_2} \varphi_{x_1}
\end{equation}
with the bare vertex
\begin{eqnarray}
\label{eq:bare_vertex_general}
&& \bar{v}_{x'_1 x'_2 x_1 x_2}\\
\nonumber &=& \delta (t'_1 - t_1) \delta (t'_2 - t_1) \delta (t_2 - t_1) (1 - \ii \eta) \bar{v}_{\alpha'_1 \alpha'_2 \alpha_1 \alpha_2}.
\end{eqnarray}
In Eq.~(\ref{eq:generating_functional_continuous}) we have employed a matrix notation similar to the one in Sec.~\ref{sec:noninteracting_generating_functional} but with multi-indices of the form $x = (\alpha, t)$ and contractions $\sum_x = \sum_\alpha \int_{- \infty}^\infty \dd t$.
In writing Eq.~(\ref{eq:bare_vertex_general}) we have dropped the infinitesimal shifts of the times at a bare vertex, compare with the time arguments in Eq.~(\ref{eq:interaction_part_of_action_continuous}).
They are made redundant by the particular choice of the equal-time value of the free propagator (\ref{eq:free_propagator_continuous}).

\subsection{Diagrammatic expansion and 1PI flow equations}

Based on the functional-integral representation (\ref{eq:generating_functional_continuous}) of the interacting generating functional, a diagrammatic expansion of the Green functions can be derived in the standard way, see appendix \ref{sec:diagrammatic_expansion}.
As usual, one can choose to work in frequency representation.
Details on the relevant Fourier transforms can be found in appendix \ref{sec:frequency_representation}.

As a next step, we introduce 1PI vertex functions and derive FRG flow equations for them.
This can be done using generating functionals, starting from the one of the Green functions.
The procedure is analogous to the one in Matsubara or Keldysh formalism \cite{Metzner12}, but for definiteness we briefly show it in appendix \ref{sec:derivation_of_flow_equations}.
The flow equation of a general 1PI $n$-particle vertex function is given by Eq.~(\ref{eq:general_flow_of_1PI_vertex_functions}).
As the first two instances ($n = 1, 2$), we obtain the flow equation of the self-energy
\begin{equation}
\label{eq:general_flow_of_self-energy}
\dot{\Sigma}^\lambda_{x'| x} = - \ii \gamma^\lambda_{x' y'| x y} S^\lambda_{y| y'}
\end{equation}
and the one of the 1PI two-particle vertex function
\begin{eqnarray}
\label{eq:general_flow_of_two-particle_vertex}
\nonumber && \dot{\gamma}^\lambda_{x' y'| x y}\\
\nonumber &=& - \ii \gamma^\lambda_{x' y' a'| x y a} S^\lambda_{a| a'}\\
&& + \ii \gamma^\lambda_{x' y'| a b} S^\lambda_{a | a'} G^\lambda_{b| b'} \gamma^\lambda_{a' b'| x y}\\
\nonumber && + \ii \gamma^\lambda_{x' b'| a y} \left( S^\lambda_{a | a'} G^\lambda_{b| b'} + S^\lambda_{b| b'} G^\lambda_{a| a'} \right) \gamma^\lambda_{a' y'| x b}\\
\nonumber && - \ii \gamma^\lambda_{y' b'| y a} \left( S^\lambda_{a| a'} G^\lambda_{b| b'} + S^\lambda_{b| b'} G^\lambda_{a| a'} \right) \gamma^\lambda_{a' x'| b x}.
\end{eqnarray}
Here, the dot above $\Sigma$ and $\gamma$ denotes the derivative with respect to $\lambda$ and $S^\lambda = G^\lambda (g^\lambda)^{-1} \dot{g}^\lambda (g^\lambda)^{-1} G^\lambda$ is the single-scale propagator.

Since the flow equation of the 1PI $n$-particle vertex function contains the 1PI $(n+1)$-particle vertex function, all of the flow equations are coupled.
In Sec.~\ref{sec:flow_equation_for_1PI_two-particle_vertex} below, we truncate this infinite hierarchy by neglecting the 1PI three-particle vertex function in Eq.~(\ref{eq:general_flow_of_two-particle_vertex}).
Due to Dyson's equation $G = 1 / (g^{-1} - \Sigma)$, one is then left with the task of solving a closed set of differential equations for the self-energy and the 1PI two-particle vertex function.
In Sec.~\ref{sec:flow_equation_for_1PI_two-particle_vertex} we also neglect the flow of the self-energy and retain only the flow equation for the two-particle vertex function.

Lastly, we note that the zero-temperature formalism can be used for slightly more general problems than to study ground-state properties.
In appendix \ref{sec:generalization_of_zero-temperature_formalism} we discuss how it can be adapted to systems in particular excited states.

\section{One-loop FRG approach to the X-ray-absorption singularity in metals}
\label{sec:FRG_approach}

In the following we devise a specific one-loop 1PI FRG approach to the model described in Sec.~\ref{sec:model} that is based on the (real-time) zero-temperature formalism.
The goal is to obtain the correct leading-logarithmic result for the X-ray absorption rate.
To achieve this, we perform approximations analogous to those of Roulet et al. \cite{Roulet69}.
We discuss why a \emph{one-loop} truncation suffices to capture the leading logarithms.
In fact, we show that the parquet-based scheme by Roulet et al. and the following one-loop FRG approach are completely equivalent.

\subsection{Cutoff and initial conditions}
\label{sec:cutoff_and_initial_conditions}

In order to introduce the flow parameter directly into the Hartree-dressed propagator rather than into the free propagator, we first absorb the deep-state Hartree self-energy $\Sigma_d^H = g \xi_0$ into the latter.
We achieve this by formally adding and subtracting a term $\Sigma_d^H a_d^\dagger a_d$ in the Hamiltonian; in terms of the action, this corresponds to adding and subtracting a term $\Sigma_d^H \bar{\varphi}^m_d \varphi^{m-1}_d$ in the square brackets in Eq.~(\ref{eq:intermediate_matrix_elements}).
The added term is then absorbed into the free action such that in the deep-state subspace the Hartree-dressed propagator given by
\begin{equation}
\label{eq:Hartree-dressed_deep-state_propagator}
G_d^H (\omega)^{-1} = G_d^0 (\omega)^{-1} - \Sigma_d^H = \omega - \ii 0^+
\end{equation}
takes on the role of the free propagator.
In Eq.~(\ref{eq:Hartree-dressed_deep-state_propagator}) the renormalized deep level has been set to $\tilde{\varepsilon}_d = 0$, see Sec.~\ref{sec:setting_renormalized_deep_level_to_zero}.
The subtracted term $\Sigma_d^H a_d^\dagger a_d$ is treated as a single-particle perturbation.
It will cancel out, see below.

We choose to employ a sharp frequency cutoff that is inserted only into the real part of the Hartree-dressed deep-state propagator,
\begin{equation}
\label{eq:introduction_of_cutoff}
G_d^{H, \lambda} (\omega) = \Theta (|\omega| - \lambda) \frac{1}{\omega} + \ii \pi \delta (\omega).
\end{equation}
No cutoff is introduced into the conduction-state propagator.
For the initial value of the flow parameter, we will consider the limit $\lambda_\text{ini} \to \infty$.
At the final value $\lambda_\text{fin} = 0$, the original model is recovered.

In the following we determine the initial values of the 1PI vertex functions, starting with the first order of the self-energy.
The deep-state Hartree diagram, which consists of a single conduction-state loop, is not affected by the cutoff.
It exactly cancels with the single-particle perturbation mentioned above: this diagram has already been accounted for in Eq.~(\ref{eq:Hartree-dressed_deep-state_propagator}).
At $\lambda_\text{ini}$ the Hartree contribution to the local conduction-electron self-energy evaluates to
\begin{subequations}
\label{eq:initial_conduction-electron_self-energy}
\begin{eqnarray}
\hspace{-4ex} \Sigma_c^{H, \lambda_\text{ini}} &=& V \Sigma_{k' k}^{H, \lambda_\text{ini}}\\
\label{eq:initial_conduction-electron_self-energy_integral} \hspace{-4ex} &=& - \ii U \int \frac{\dd \omega}{2 \pi} e^{\ii \omega \eta'} \left[ \frac{\Theta (|\omega| - \lambda_\text{ini})}{\omega} + \ii \pi \delta (\omega) \right]\\
\hspace{-4ex} &=& U,
\end{eqnarray}
\end{subequations}
which is the same as without cutoff and which cancels the single-particle perturbation that arises when the interaction term in Eq.~(\ref{eq:Hamiltonian}) is brought into the standard form.
In Eq.~(\ref{eq:initial_conduction-electron_self-energy_integral}) a convergence factor $e^{\ii \omega \eta'}$ with $\eta' \to 0^+$ has been included as part of the Hartree-dressed deep-state propagator, see Eq.~(\ref{eq:frequency_representation_of_free_propagator_diagonal}).
Here, it is important to take the limit $\eta' \to 0^+$ before the limit $\lambda_\text{ini} \to \infty$.

Let us now consider an arbitrary 1PI diagram with at least two vertices and conclude that its initial value is negligible.
If it contains a deep-state propagator that connects two different vertices, this propagator is replaced by its delta-function part at $\lambda_\text{ini}$.
The result is negligible in leading logarithmic order according to Roulet et al.~\cite{Roulet69}.
If the diagram does not contain any such deep-state propagator, then all its external legs are deep-state ones.
But a 1PI subdiagram of this form does not enter the parquet diagrams that contain the leading logarithms, see Sec.~\ref{sec:logarithmic_divergences}.
In the leading-logarithmic approximation, the initial conditions are thus fully determined by diagrams with just a single vertex.
Consequently, the initial value of the 1PI two-particle vertex is given by the bare vertex,
\begin{equation}
\hat{\gamma}_{\lambda_\text{ini}} (\omega, \omega'; \Omega) = U = \bar{\gamma}_{\lambda_\text{ini}} (\omega, \omega'; X),
\end{equation}
while the initial values of all other 1PI vertex functions, including the self-energy, vanish.

\subsection{Flow equation for $\gamma$}
\label{sec:flow_equation_for_1PI_two-particle_vertex}

In order to truncate and solve the set of FRG flow equations, we neglect the flow of the self-energy and of the 1PI three-particle vertex so that the values of both remain zero.
This means that we set the right-hand side of Eq.~(\ref{eq:general_flow_of_self-energy}) to zero and neglect the first addend on the right-hand side of Eq.~(\ref{eq:general_flow_of_two-particle_vertex}).
That we can indeed renounce corrections to the self-energy for the leading-logarithmic approximation is evident from the diagrammatic discussion in Ref.~\cite{Roulet69}, see also our Sec.~\ref{sec:single-particle_Green_function}.

\begin{figure}
\includegraphics[scale=0.75]{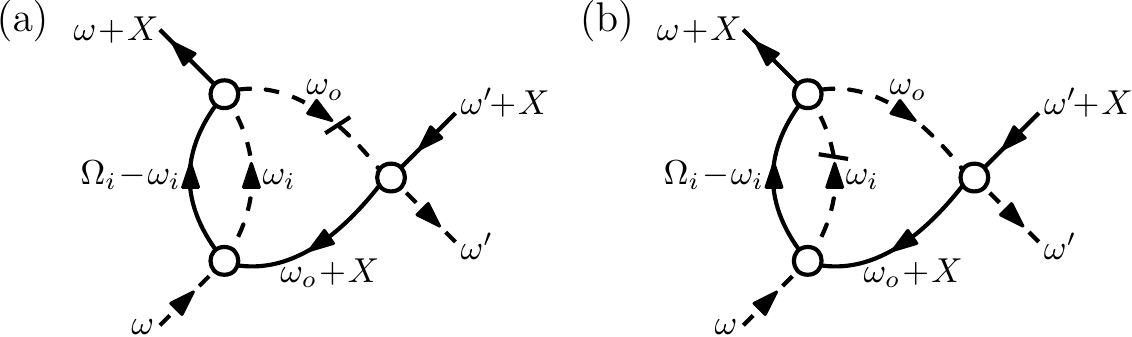}
\caption{\label{fig:derivative_of_crossed_bubble}
The diagrammatic contributions to $\dd \gamma_\lambda / \dd \lambda$ that result from taking the derivative of the diagram in Fig.~\ref{fig:crossed_bubble}.
Again, $\Omega_i$ stands for $\omega_o + \omega + X$.
In each diagram the crossed-out dashed line represents the differentiated deep-state propagator.
(a) A leading contribution. Removing the crossed-out line, which is in the outer bubble, would render the diagram one-particle reducible.
(b) A subleading contribution. Removing the crossed-out line, which is in the inner bubble, would leave the diagram one-particle irreducible.}
\end{figure}

Let us clarify why the 1PI three-particle vertex does not affect the leading logarithmic order of the 1PI two-particle vertex via the flow equations.
The argument is based on the properties of individual diagrams.
In the diagrammatic derivation \cite{Jakobs07} of the flow equation for the 1PI two-particle vertex $\gamma_\lambda$, the derivative with respect to the flow parameter acts on each of the contributing diagrams.
In each diagram, according to the product rule, every deep-state propagator needs to be differentiated.
Therefore, $\dd \gamma_\lambda / \dd \lambda$ is represented by a sum of diagrams in each of which the derivative acts on some particular dashed line; examples for this type of diagram are shown in Fig.~\ref{fig:derivative_of_crossed_bubble}.
Let us now consider any such diagram that results from differentiating one of the parquet diagrams as those contain all of the important contributions.
Because of Eq.~(\ref{eq:introduction_of_cutoff}) the frequency of the differentiated propagator satisfies $|\omega| = \lambda$.
With regard to the real parts of the deep-state propagators, this frequency has the smallest absolute value of all deep-state frequencies.
As far as the leading logarithms are concerned, the differentiated propagator then has to be in one of the outermost bubbles; this follows from the discussion at the end of Sec.~\ref{sec:third-order_example} about the integration regions that give rise to the leading logarithms.
Removing this propagator would render the diagram one-particle \emph{reducible}.
In contrast all diagrams contributing to $\dd \gamma_\lambda / \dd \lambda$ that stem from the 1PI three-particle vertex, i.e., that represent the first term in Eq.~(\ref{eq:general_flow_of_two-particle_vertex}), would remain one-particle \emph{irreducible} if the respective differentiated propagator was removed.
Consequently, the leading logarithmic contributions cannot originate from the 1PI three-particle vertex so that it can indeed be neglected.
This shows that, for a sharp frequency cutoff in the deep-state propagator, a \emph{one-loop} truncation already captures all important contributions of the parquet diagrams even though it does not account for the exact values of these diagrams.

As an example we illustrate the above argument for the third-order parquet diagram in Fig.~\ref{fig:crossed_bubble}.
Its derivative with respect to $\lambda$ is the sum of the two diagrams shown in Fig.~\ref{fig:derivative_of_crossed_bubble}.
If we removed the respective differentiated propagator, diagram (a) would become one-particle reducible, whereas diagram (b) would remain one-particle irreducible, i.e., the latter stems from the 1PI three-particle vertex.
In diagram (a) the frequencies (of the real parts of the deep-state propagators) satisfy $|\omega_o| = \lambda \le |\omega_i|$ and in diagram (b) they satisfy $|\omega_o| \ge \lambda = |\omega_i|$.
Since we have shown in Sec.~\ref{sec:third-order_example} for the diagram in Fig.~\ref{fig:crossed_bubble} that only the integration region $|\omega_o| < |\omega_i|$ gives rise to the leading logarithms, the contribution to the flow represented by diagram (b) is negligible.

Following Roulet et al.~\cite{Roulet69}, we approximate the propagators by neglecting the real part of the conduction-state propagator and the imaginary part of the deep-state propagator because they do not give rise to the logarithmic divergence in a bubble, cf. Eq.~(\ref{eq:origin_of_logarithm}).
This step can be performed only after the evaluation of Eq.~(\ref{eq:initial_conduction-electron_self-energy}) for the initial conditions, where the imaginary part of the deep-state propagator contributes half of the result.
Within these approximations the local conduction-electron propagator does not depend on the flow parameter and is given by
\begin{equation}
G_c (\omega) = - \ii \pi \frac{\rho}{V} \, \sgn (\omega) \Theta (\xi_0 - |\omega|)
\end{equation}
and the Hartree-dressed single-scale propagator assumes the form
\begin{equation}
\label{eq:single-scale_propagator}
S_d^\lambda (\omega) = \frac{\dd}{\dd \lambda} G^{H, \lambda}_d (\omega) = - \frac{\delta (|\omega| - \lambda)}{\omega}.
\end{equation}

\begin{figure}
\includegraphics[scale=0.75]{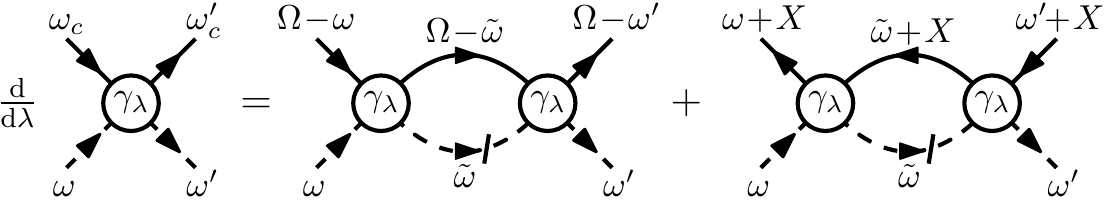}
\caption{\label{fig:flow_of_full_vertex}
Diagrammatic representation of the flow equation (\ref{eq:flow_of_full_vertex}) for the 1PI two-particle vertex.
The crossed-out dashed line stands for the single-scale propagator $S_d^\lambda$.
The first term represents the particle-particle channel while the second term represents the (exchange) particle-hole channel.
The external frequencies are related via $\omega_c = \Omega - \omega = \omega' + X$ and $\omega'_c = \Omega - \omega' = \omega + X$.}
\end{figure}

There remains to be solved the flow equation for the 1PI two-particle vertex.
While its general form is as stated in Eq.~(\ref{eq:general_flow_of_two-particle_vertex}), it now assumes the closed form
\begin{subequations}
\label{eq:flow_of_full_vertex}
\begin{eqnarray}
\nonumber \! \! && \frac{\dd}{\dd \lambda} \hat{\gamma}_\lambda (\omega, \omega'; \Omega)\\
\! \! &=& \frac{\dd}{\dd \lambda} \bar{\gamma}_\lambda (\omega, \omega'; X)\\
\label{eq:flow_of_full_vertex_rhs}
\! \! &=& - \frac{1}{2} \frac{\rho}{V} \int \dd \tilde{\omega} \, \delta (|\tilde{\omega}| - \lambda) \frac{1}{\tilde{\omega}}\\
\nonumber \! \! && \times \big[ \hat{\gamma}_\lambda (\omega, \tilde{\omega}; \Omega) \hat{\gamma}_\lambda (\tilde{\omega}, \omega'; \Omega) \sgn (\Omega - \tilde{\omega}) \Theta (\xi_0 - |\Omega - \tilde{\omega}|)\\
\nonumber \! \! && + \bar{\gamma}_\lambda (\omega, \tilde{\omega}; X) \bar{\gamma}_\lambda (\tilde{\omega}, \omega'; X) \sgn (\tilde{\omega} + X) \Theta (\xi_0 - |\tilde{\omega} + \! X|) \big],
\end{eqnarray}
\end{subequations}
where the frequency arguments are related via $\Omega - X = \omega + \omega'$.
The diagrammatic representation of this equation is shown in Fig.~\ref{fig:flow_of_full_vertex}.
In writing Eq.~(\ref{eq:flow_of_full_vertex}) we consider the flow only of $\gamma_{d c| d c}^\lambda$ because it is the sole part of the 1PI vertex needed to calculate the particle-hole susceptibility (\ref{eq:ph-susceptibility_with_1PI_vertex}).
We do not consider the flow of $\gamma_{c c| c c}^\lambda$ because it does not influence the flow of $\gamma_{d c| d c}^\lambda$: the single-scale propagator, which has only deep-state indices, cannot be attached to $\gamma_{c c| c c}^\lambda$.
We neither consider the flow of $\gamma_{d d| d d}^\lambda$.
Its flow equation and its contribution to the flow of $\gamma_{d c| d c}^\lambda$ both involve a bubble with two deep-state propagators.
Consequently, its influence is subleading.
This reflects that the parquet diagrams containing the leading logarithms do not comprise 1PI subdiagrams with deep-state external indices only, as mentioned in the discussion of the initial conditions close to the end of Sec.~\ref{sec:cutoff_and_initial_conditions}.
As a result of neglecting $\gamma_{d d| d d}^\lambda$, the direct particle-hole channel is absent in Eq.~(\ref{eq:flow_of_full_vertex}).

We perform a channel decomposition by defining $\dd \hat{\gamma}^\text{pp}_\lambda (\omega, \omega'; \Omega) / \dd \lambda$ as the first addend in Eq.~(\ref{eq:flow_of_full_vertex_rhs}) and $\dd \bar{\gamma}^\text{ph}_\lambda (\omega, \omega'; X) / \dd \lambda$ as the second addend.
For the choice $\hat{\gamma}^\text{pp}_{\lambda_\text{ini}} (\omega, \omega'; \Omega) = 0 = \bar{\gamma}^\text{ph}_{\lambda_\text{ini}} (\omega, \omega'; X)$, a formal integration of the flow equation leads to the decomposition of the 1PI two-particle vertex
\begin{eqnarray}
\hat{\gamma}_\lambda (\omega, \omega'; \Omega) &=& \bar{\gamma}_\lambda (\omega, \omega'; X)\\
\nonumber &=& U + \hat{\gamma}^\text{pp}_\lambda (\omega, \omega'; \Omega) + \bar{\gamma}^\text{ph}_\lambda (\omega, \omega'; X),
\end{eqnarray}
where $\Omega - X = \omega + \omega'$.
Equation~(\ref{eq:flow_of_full_vertex}) can then be rewritten in terms of the two coupled flow equations
\begin{subequations}
\label{eq:decomposed_flow_equation}
\begin{eqnarray}
\nonumber && \frac{\dd}{\dd \lambda} \hat{\gamma}^\text{pp}_\lambda (\omega, \omega'; \Omega)\\
\label{eq:decomposed_flow_equation_pp} &=& - \frac{1}{2} \frac{\rho}{V} \sum_{\tilde{\omega} = \pm \lambda} \frac{1}{\tilde{\omega}} \sgn (\Omega - \tilde{\omega}) \Theta (\xi_0 - |\Omega - \tilde{\omega}|)\\
\nonumber && \times \left[ U + \hat{\gamma}^\text{pp}_\lambda (\omega, \tilde{\omega}; \Omega) + \bar{\gamma}^\text{ph}_\lambda (\omega, \tilde{\omega}; \Omega - \omega - \tilde{\omega}) \right]\\
\nonumber && \times \left[ U + \hat{\gamma}^\text{pp}_\lambda (\tilde{\omega}, \omega'; \Omega) + \bar{\gamma}^\text{ph}_\lambda (\tilde{\omega}, \omega'; \Omega - \tilde{\omega} - \omega') \right]
\end{eqnarray}
and
\begin{eqnarray}
\nonumber && \frac{\dd}{\dd \lambda} \bar{\gamma}^\text{ph}_\lambda (\omega, \omega'; X)\\
\label{eq:decomposed_flow_equation_ph} &=& - \frac{1}{2} \frac{\rho}{V} \sum_{\tilde{\omega} = \pm \lambda} \frac{1}{\tilde{\omega}} \sgn (\tilde{\omega} + X) \Theta (\xi_0 - |\tilde{\omega} + X|)\\
\nonumber && \times \left[ U + \hat{\gamma}^\text{pp}_\lambda (\omega, \tilde{\omega}; \omega + \tilde{\omega} + X) + \bar{\gamma}^\text{ph}_\lambda (\omega, \tilde{\omega}; X) \right]\\
\nonumber && \times \left[ U + \hat{\gamma}^\text{pp}_\lambda (\tilde{\omega}, \omega'; \tilde{\omega} + \omega' + X) + \bar{\gamma}^\text{ph}_\lambda (\tilde{\omega}, \omega'; X) \right].
\end{eqnarray}
\end{subequations}
These are the contributions to the flow in the particle-particle and particle-hole channel, respectively.
In the diagrammatic language, Eq.~(\ref{eq:decomposed_flow_equation_pp}) gives rise to diagrams that can be disconnected by cutting two parallel lines, whereas the diagrams resulting from Eq.~(\ref{eq:decomposed_flow_equation_ph}) can be disconnected by cutting two antiparallel lines.
Since a diagram of one type can appear as a subdiagram in diagrams of the other type, the two differential equations are coupled.
For the contribution from each channel, we have employed the notation that features the respective natural frequency, see Eq.~(\ref{eq:frequency_notation}) and (\ref{eq:bare_bubbles}).

We now make an assumption that we will later show to be correct within logarithmic accuracy based on a self-consistency argument:
We assume that the relations
\begin{subequations}
\label{eq:assumed_relations}
\begin{eqnarray}
\hat{\gamma}^\text{pp}_\lambda (\omega, \omega'; \Omega) &=& \hat{\gamma}^\text{pp}_\lambda (|\omega|, |\omega'|; \max \{ \lambda, |\Omega| \} )\\
\bar{\gamma}^\text{ph}_\lambda (\omega, \omega'; X) &=& \bar{\gamma}^\text{ph}_\lambda (|\omega|, |\omega'|; \max \{ \lambda, |X| \} )
\end{eqnarray}
\end{subequations}
hold, which are trivially satisfied at the start of the flow.
We use these relations to rewrite the vertex functions appearing on the right-hand side of the flow equations (\ref{eq:decomposed_flow_equation}).

In the terms representing the cross feedback, we subsequently approximate the third frequency argument.
For example, in the term $\hat{\gamma}^\text{pp}_\lambda (\omega, \pm \lambda; \omega \pm \lambda + X) = \hat{\gamma}^\text{pp}_\lambda (|\omega|, \lambda; \max \{ \lambda, |\omega \pm \lambda + X| \} )$ appearing in Eq.~(\ref{eq:decomposed_flow_equation_ph}), we approximate
\begin{equation}
\label{eq:max_approximation}
\max \{ \lambda, |\omega \pm \lambda + X| \} \approx \max \{ \lambda, |\omega| \}.
\end{equation}
This step is analogous to neglecting the second addend in Eq.~(\ref{eq:internal_bubble_break_down}).
It can be justified as follows.
For $\lambda \ge |X|$ the approximation (\ref{eq:max_approximation}) is correct within a factor of three.
Such a factor is negligible because, based on the considerations of Roulet et al.~\cite{Roulet69}, we expect $\hat{\gamma}^\text{pp}_\lambda$ to be a slowly varying function of its arguments; this expectation will be confirmed by the final result.
For $\lambda < |X|$ the two summands for $\tilde{\omega} = \pm \lambda$ cancel each other at least to a large extent because the factor $\sgn (\pm \lambda + X)$ does not cancel the sign of $1 / (\pm \lambda)$ anymore.
Consequently, the final part with $\lambda < |X|$ of the flow in the particle-hole channel does not contribute to building the leading logarithms.
This corresponds to the observation that small frequencies with $|\tilde{\omega}| < |X|$ do not contribute to the logarithmic divergence of the bare particle-hole bubble, see Eq.~(\ref{eq:origin_of_logarithm}).
For the other cross-feedback terms in the flow equations, we apply approximations analogous to Eq.~(\ref{eq:max_approximation}).
The justification is similar.

Next, we replace the step functions $\Theta (\xi_0 - |\Omega \mp \lambda|)$ and $\Theta (\xi_0 - |\pm \lambda + X|)$ occurring in Eq.~(\ref{eq:decomposed_flow_equation}) with $\Theta (\xi_0 - \lambda)$.
When compared with the parquet-based scheme by Roulet et al.~\cite{Roulet69}, this corresponds to neglecting the third addend in Eq.~(\ref{eq:internal_bubble_break_down}) and to replacing the integration boundaries by $\pm \xi_0$ in Eq.~(\ref{eq:origin_of_logarithm}) and (\ref{eq:crossed_bubble_result}).
To motivate this approximation, consider integrating the flow equations by applying $- \int_0^{\lambda_\text{ini}} \dd \lambda \ldots$
The resulting $\lambda$-integrals take on the role of the frequency integral in a bubble.
Provided that $|\Omega| \ll \xi_0$ or $|X| \ll \xi_0$, respectively, the replacement above is wrong only for certain $\lambda \approx \xi_0$, but the leading contribution that builds the logarithm comes from $\lambda \ll \xi_0$.
Indeed, said conditions are satisfied: For the particle-hole susceptibility (\ref{eq:ph-susceptibility_with_1PI_vertex}) near threshold, the values of $\hat{\gamma}^\text{pp}_{\lambda = 0} (\omega, \omega'; \omega + \omega' + \nu)$ and $\bar{\gamma}^\text{ph}_{\lambda = 0} (\omega, \omega'; \nu)$ are important only for $|\omega|, |\omega'|, |\nu| \ll \xi_0$; it follows that the values of $\hat{\gamma}^\text{pp}_\lambda$ and $\bar{\gamma}^\text{ph}_\lambda$ are relevant only with all frequency arguments being small -- even for the cross-feedback terms in Eq.~(\ref{eq:decomposed_flow_equation}), compare the discussion in the paragraphs following Eq.~(\ref{eq:crossed_bubble_result}).
Consequently, the error made by replacing the step functions is negligible.
Due to the factors $\Theta (\xi_0 - \lambda)$, the actual flow now starts at $\lambda = \xi_0$.
This constitutes our last approximation.

Since the vertex functions occurring in the flow equations do not depend on the sign of $\tilde{\omega} = \pm \lambda$ anymore, we can easily perform the sums over $\tilde{\omega}$ in Eq.~(\ref{eq:decomposed_flow_equation}), e.g.,
\begin{equation}
\sum_{\tilde{\omega} = \pm \lambda} \frac{1}{\tilde{\omega}} \sgn (\tilde{\omega} + X) = \frac{2}{\lambda} \Theta (\lambda - |X|).
\end{equation}
A formal integration with respect to the flow parameter, starting from $\lambda_\text{ini}$ down to some value $\lambda$, then leads to
\begin{subequations}
\label{eq:integrated_flow}
\begin{eqnarray}
\nonumber && \hat{\gamma}^\text{pp}_\lambda (\omega, \omega'; \Omega)\\
\label{eq:integrated_flow_pp} &=& - \frac{\rho}{V} \int_{\max \{ \lambda, |\Omega| \} }^{\xi_0} \frac{\dd \lambda'}{\lambda'}\\
\nonumber && \times \left[ U + \hat{\gamma}^\text{pp}_{\lambda'} (|\omega|, \lambda'; \lambda') + \bar{\gamma}^\text{ph}_{\lambda'} (|\omega|, \lambda'; \max \{ \lambda', |\omega| \} ) \right]\\
\nonumber && \times \left[ U + \hat{\gamma}^\text{pp}_{\lambda'} (\lambda', |\omega'|; \lambda') + \bar{\gamma}^\text{ph}_{\lambda'} (\lambda', |\omega'|; \max \{ \lambda', |\omega'| \} ) \right]
\end{eqnarray}
and
\begin{eqnarray}
\nonumber && \bar{\gamma}^\text{ph}_\lambda (\omega, \omega'; X)\\
&=& \frac{\rho}{V} \int_{\max \{ \lambda, |X| \} }^{\xi_0} \frac{\dd \lambda'}{\lambda'}\\
\nonumber && \times \left[ U + \hat{\gamma}^\text{pp}_{\lambda'} (|\omega|, \lambda'; \max \{ \lambda', |\omega| \} ) + \bar{\gamma}^\text{ph}_{\lambda'} (|\omega|, \lambda'; \lambda') \right]\\
\nonumber && \times \left[ U + \hat{\gamma}^\text{pp}_{\lambda'} (\lambda', |\omega'|; \max \{ \lambda', |\omega'| \} ) + \bar{\gamma}^\text{ph}_{\lambda'} (\lambda', |\omega'|; \lambda') \right]
\end{eqnarray}
\end{subequations}
for $\xi_0 > \lambda, |\Omega|, |X|$.
As claimed above, the relations (\ref{eq:assumed_relations}) follow from these flow equations and are thus validated within logarithmic accuracy: On the right-hand side of Eq.~(\ref{eq:integrated_flow_pp}), only the absolute values of $\omega$ and $\omega'$ enter and only $\max \{ \lambda, |\Omega| \}$ appears with no separate dependence on $\lambda$ or $\Omega$; the analogue holds for the particle-hole channel.
For this reason we can from now on even write $\hat{\gamma}^\text{pp}_0$ and $\bar{\gamma}^\text{ph}_0$ instead of $\hat{\gamma}^\text{pp}_{\lambda'}$ and $\bar{\gamma}^\text{ph}_{\lambda'}$ in the integrands in Eq.~(\ref{eq:integrated_flow}).
Furthermore, the flow equations (\ref{eq:integrated_flow}), in conjunction with the vanishing initial conditions, imply that $\hat{\gamma}^\text{pp}_\lambda (\omega, \omega'; \Omega)$ does not depend on $\omega^{(\prime)}$ if $|\omega^{(\prime)}| \le |\Omega|$ and the same for $\bar{\gamma}^\text{ph}_\lambda (\omega, \omega'; X)$ if $|\omega^{(\prime)}| \le |X|$.
It is therefore reasonable to introduce the shorthand notation
\begin{subequations}
\begin{eqnarray}
\hat{\gamma}^\text{pp}_\lambda (\Omega) &=& \hat{\gamma}^\text{pp}_\lambda (\omega, \omega'; \Omega) \qquad \text{if } |\omega|, |\omega'| \le |\Omega|\\
\bar{\gamma}^\text{ph}_\lambda (X) &=& \bar{\gamma}^\text{ph}_\lambda (\omega, \omega'; X) \qquad \text{if } |\omega|, |\omega'| \le |X|.
\end{eqnarray}
\end{subequations}
With this the integrated flow equations at $\lambda_\text{fin} = 0$ assume the form
\begin{subequations}
\label{eq:final_flow}
\begin{eqnarray}
&& \hat{\gamma}^\text{pp}_0 (\omega, \omega'; \Omega)\\
\nonumber &=& - \frac{\rho}{V} \int_{|\Omega|}^{\xi_0} \frac{\dd \lambda}{\lambda} \left[ U + \hat{\gamma}^\text{pp}_0 (|\omega|, \lambda; \lambda) + \bar{\gamma}^\text{ph}_0 (\max \{ \lambda, |\omega| \} ) \right]\\
\nonumber && \hphantom{- \frac{\rho}{V} \int_{|\Omega|}^{\xi_0}} \times \left[ U + \hat{\gamma}^\text{pp}_0 (\lambda, |\omega'|; \lambda) + \bar{\gamma}^\text{ph}_0 (\max \{ \lambda, |\omega'| \} ) \right]
\end{eqnarray}
and
\begin{eqnarray}
&& \bar{\gamma}^\text{ph}_0 (\omega, \omega'; X)\\
\nonumber &=& \frac{\rho}{V} \int_{|X|}^{\xi_0} \frac{\dd \lambda}{\lambda} \left[ U + \hat{\gamma}^\text{pp}_0 (\max \{ \lambda, |\omega| \} ) + \bar{\gamma}^\text{ph}_0 (|\omega|, \lambda; \lambda) \right]\\
\nonumber && \hphantom{\frac{\rho}{V} \int_{|X|}^{\xi_0}} \times \left[ U + \hat{\gamma}^\text{pp}_0 (\max \{ \lambda, |\omega'| \} ) + \bar{\gamma}^\text{ph}_0 (\lambda, |\omega'|; \lambda) \right].
\end{eqnarray}
\end{subequations}

\subsection{Relation to parquet approach of Roulet et al.}

The integrated flow equations (\ref{eq:final_flow}) are identical to Eq.~(A1) and (A2) of Roulet et al.~\cite{Roulet69}.
There are differences only in the notation, in particular the authors of Ref.~\cite{Roulet69} introduced logarithmic variables for all frequencies.
They solved these integral equations without further approximations and used the resulting 1PI vertex to determine $\im \chi (\nu)$.
We can without changes adopt the steps of Roulet et al. to determine $\im \chi (\nu) = - \pi (\rho / V) (\xi_0 / \nu)^{2g} \Theta (\nu)$, which provides the shape of the absorption rate near the threshold via $R(\nu) = - 2 |W|^2 \im \chi (\nu)$.
We refrain from repeating these steps here.
We have thus established that the one-loop FRG approach presented in this section leads to the exact same result for the X-ray absorption rate as the parquet-based scheme by Roulet et al.
In particular, this proves that the one-loop FRG approach captures all leading logarithms.

On top of that, we argue that the two approaches do not only produce the identical result but are fully equivalent on a detailed level.
In fact, the various approximation steps in the two approaches can be identified with each other.
The first approximation performed by Roulet et al. in Ref.~\cite{Roulet69} is to replace the totally irreducible interaction $R$ by the bare interaction.
This reduces the diagrams under consideration to the parquet diagrams in the particle-particle and exchange particle-hole channel.
The same reduction results in the FRG approach from neglecting the three-particle vertex and $\gamma_{d d| d d}$.
(The role of neglecting $\gamma_{d d| d d}$ is to eliminate the direct particle-hole channel; in the treatment by Roulet et al., this channel is embedded in the irreducible interaction $R$ and then neglected.)
Note that disregarding the three-particle vertex in the FRG approach brings about one additional approximation, namely that the internal frequency integrations in parquet diagrams with crossed channels are performed only partly.
In the approach of Roulet et al., the same restriction in the frequency integrations is a by-product of the logarithmic approximation, see below.

Reference \cite{Roulet69} continues with some approximations which we transferred one-to-one to our FRG approach: neglecting the real part of $G_c$, neglecting the deep-state self-energy except for a static contribution, and neglecting the imaginary part of $G_d$.
The next step is the "logarithmic approximation" ten lines above Eq.~(29) of Ref.~\cite{Roulet69}.
It has its direct counterpart in our Eq.~(\ref{eq:max_approximation}).
Furthermore, approximating the upper integration boundary by $\xi_0$ on the right-hand side of Eq.~(29) in Ref.~\cite{Roulet69} corresponds to replacing $\Theta(\xi_0 - |\Omega \mp \lambda|)$ by $\Theta(\xi_0 - \lambda)$ in our Sec.~\ref{sec:flow_equation_for_1PI_two-particle_vertex}.
Finally, the abovementioned restriction in the internal frequency integrations follows in Ref.~\cite{Roulet69} from the logarithmic approximation when Eq.~(29) and (31) of that reference are combined.
The restriction is fully realized in Eq.~(34) of Ref.~\cite{Roulet69}, where the argument of $I_1$ in the first integrand is not greater than the integration variable $t_i$.
When the inner bubble contained in $I_1$ is evaluated according to the second integral, this argument takes on the role of $\beta$ which is the upper bound of the second integral.
The integration variable of the outer bubble, i.e., of the first integral, is therefore greater than the one of the inner bubble.
For the corresponding frequencies of the bubbles follows the reverse relation, and this is precisely the same restriction as the one established by the one-loop FRG.

In the parquet approach of Ref.~\cite{Roulet69}, it remains to bring the equations into a solvable form.
To achieve this Roulet et al. invoke a ``trick by Abrikosov and Sudakov'', see p.~1081 and the appendix of Ref.~\cite{Roulet69}.
This step at the very end of their solution can indeed be identified with the introduction of a sharp frequency cutoff at the very beginning of the FRG treatment: When applying this trick, one considers the general structure of a parquet diagram reducible in a given channel; one identifies among the outermost bubbles the one with the smallest absolute value of the deep-state frequency; to both sides of this bubble, there are full 1PI vertices that are restricted to contain only greater deep-state frequencies; finally, the abovementioned smallest frequency is integrated over.
The concept of a smallest deep-state frequency which is integrated over corresponds precisely to a sharp frequency cutoff in the deep-state propagator and a formal integration of the FRG flow equations.
In this way Fig.~13 of Ref.~\cite{Roulet69} (which shows only one channel, has wrongly directed arrows on the deep-state lines, and does not formally add the kernels $I_1$ to vertex functions $\gamma$ on the left and on the right) anticipates the graphical representation of the FRG flow equation in our Fig.~\ref{fig:flow_of_full_vertex}.

We thus have established the full equivalence of both approaches.
The only difference lies in the order of the steps.
Our FRG approach starts by introducing a cutoff and continues with approximations to the flow equations.
Roulet et al., on the other hand, apply equivalent approximations to the parquet equations and use the cutoff only at the very end to rewrite and solve the resulting equations.

We note that our particular choice of the cutoff is crucial for the equivalence discussed above.
For ill-conceived cutoffs the 1PI three-particle vertex could significantly influence the flow of the 1PI two-particle vertex such that a one-loop truncation would not capture the leading logarithms.
However, we expect a one-loop truncation to be sufficient in the case of any sensible cutoff that regularizes the divergences of the bare bubbles during the entire flow.

\section{Conclusion and outlook}
\label{sec:conclusion}

Historically, the concept of summing up all parquet diagrams with bare lines was developed to construct the leading-logarithmic approximation for models in which bubbles in different channels produce simple logarithmic divergences \cite{Diatlov57, Abrikosov65, Bychkov66, Roulet69}.
In a paradigmatic case, Roulet et al. derived from this approach the leading approximation for the rate of X-ray absorption in metals close to the threshold frequency \cite{Roulet69}.
In the present paper, we have shown that a standard one-loop FRG approximation with sharp frequency cutoff reproduces identically the parquet-based leading-logarithmic approximation of Roulet et al.
There is a detailed correspondence between the approaches; in particular, the ``trick by Abrikosov and Sudakov'' to evaluate the approximate parquet equations corresponds to the introduction of the cutoff in the FRG.
In total, the two approaches can be understood as different viewpoints on the same technical steps.
Extending our one-loop scheme to a multiloop scheme on the analogy of Ref.~\cite{Kugler18a} would result in a change on the subleading level without leading to a controlled improvement.
We explained why the parts of the parquet diagrams that are not captured by the one-loop FRG (and not by the treatment of Roulet et al.) are subleading.
The traditional understanding that low-order RG approximations can reproduce the leading-logarithmic parquet result for models with simple logarithmic divergences of the bubbles \cite{Anderson70, Abrikosov70, Fowler71, Solyom74, Solyom74a, Kimura73, Solyom79, Irkhin01, Binz03} is thus reconfirmed also in this case.

For the whole class of these mostly zero- and one-dimensional models, we therefore do not expect the multiloop FRG to be advantageous on the leading-logarithmic level.
There remains at least the benefit that the multiloop scheme provides a leading-logarithmic approximation for any choice of flow parameter, as long as convergence is reached.
This establishes more flexibility compared to our analysis of the one-loop scheme, which deals with a specific cutoff.
While we expect that our analysis can be transferred to other regularizing cutoffs, this should be reexamined in each case.
In fact, the effort to do so would be worthwhile since the leading-logarithmic properties can often be extracted analytically from one-loop RG (or from the parquet equations) as seen in this study and in almost all corresponding references mentioned in the Introduction.
The multiloop scheme, in contrast, provides only a numerical solution.

Consequently, we expect that the multiloop scheme will mainly be found useful to evaluate the full parquet approximation, which is typically used for two-dimensional models.
The resulting approximation benefits from preserving certain sum rules and conservation laws and the Mermin-Wagner theorem.
Results of multiloop investigations of the two-dimensional Hubbard model are promising in this respect \cite{Tagliavini19, Hille20a, Hille20}.
As the multiloop FRG is not restricted to approximate the totally irreducible vertex by the bare one, it might also turn out to be helpful for constructing diagrammatic extensions of the dynamical mean-field theory.

In the present paper, we established a formulation of the FRG within the (real-time) zero-temperature formalism.
This formalism is more restrictive than the Matsubara and Keldysh formalisms because it only provides access to ground-state properties.
Additionally, its application requires that the noninteracting ground state is not, as a matter of different symmetries, orthogonal to the interacting one \cite{Negele88}.
However, it has the advantage that it is based on real times or frequencies and therefore does not require an analytic continuation from the imaginary to the real frequency axis.
Such an analytic continuation is a significant complication for numerical FRG results obtained within Matsubara formalism, see, e.g., Ref.~\cite{Karrasch08}.
Compared to Keldysh formalism, the zero-temperature formalism is easier to work with as it involves only a single time axis instead of a two-branch time contour.
The two branches in Keldysh formalism give rise to different components (say, chronological, lesser, greater, antichronological) of Green functions, whereas there is only a single component in the zero-temperature formalism.
Due to these features, we expect that the zero-temperature FRG developed in this paper can have useful future applications.

Several topics for future research naturally arise from the considerations set forth in this paper.
First, it should be clarified how our observations made within the real-time zero-temperature formalism can be transferred to formulations of the FRG within the Matsubara formalism.
This is important to achieve a more detailed comparability to the works of Kugler and von Delft on the X-ray-absorption problem \cite{Kugler18, Kugler18a}, which use the Matsubara formalism.
In particular, the nature of the improvements achieved by multiloop FRG as reported in Ref.~\cite{Kugler18a} could be clarified: either the one-loop schemes used in that reference are suboptimal in that they miss some leading logarithms or the observed changes due to the multiloop scheme are in the uncontrolled regime.
This question is still open as Ref.~\cite{Kugler18a} disregards the subleading difference between the exact sum of the parquet diagrams and the leading-logarithmic solution of Roulet et al. from Ref.~\cite{Roulet69}.
The transfer of our observations to the Matsubara formalism is important also from a more general perspective due to the widespread use of the Matsubara FRG as a tool to investigate low-dimensional fermionic systems \cite{Metzner12, Kopietz10, Platt13}.
We started investigations of the leading-logarithmic approximation to the X-ray-absorption rate using Matsubara FRG.
They indicate that the central message of this paper can indeed be transferred to the Matsubara case: a reasonably crafted one-loop Matsubara FRG scheme reproduces the leading-logarithmic approximation.
We observe that passing over to continuous Matsubara frequencies at zero temperature and setting $\tilde{\varepsilon}_d$ to zero requires particular care within Matsubara formalism.
We intend to address these points in a future publication.

Another topic for future research is the mechanism by which the one-loop FRG captures the leading logarithms.
The corresponding reasoning in Sec.~\ref{sec:flow_equation_for_1PI_two-particle_vertex} was based on individual diagrams; this allowed us to stress the close analogy to the leading-logarithmic parquet approximation.
We expect, however, that an argument based completely on the structure of the flow equations could be more efficient.
This could help with another task, namely, to construct FRG approximations that treat subleading contributions consistently.
Whether the multiloop FRG with dressed propagators can contribute to achieve the latter goal remains to be clarified as well.

Furthermore, it is desirable to extend the considerations of this paper to nonequilibrium situations which can be described within the Keldysh formalism.
This would allow for interesting applications to model systems for quantum dots and wires.
For example, one could expand on the FRG study of nonequilibrium Kondo physics in Ref.~\cite{Schmidt10}, which does not discuss the question of a consistent treatment of logarithmic divergences.
This could make it possible to address open questions concerning the influence of a magnetic field and to achieve a methodological comparison with the real-time RG approach to nonequilibrium Kondo physics of Ref.~\cite{Reininghaus14}.

\begin{acknowledgments}
We are grateful to Fabian Kugler and Jan von Delft for raising our interest in the topic of this work and for useful discussions.
We are obliged to Andrey Katanin for instructive explanations on the role of logarithmic divergences in two-dimensional systems.
We thank Volker Meden for many stimulating discussions in the broader context of this work and for a critical reading of the manuscript.
This work was supported by the Deutsche Forschungsgemeinschaft via RTG 1995.
\end{acknowledgments}

\appendix

\section{Details on the continuous functional-integral notation}
\label{sec:details_on_continuum_limit}

In this appendix we discuss the continuous form which the parts of the action acquire in the limit $M \to \infty$.

We can cast the free action (\ref{eq:free_part_of_action_discrete}) into either of the following forms
\begin{subequations}
\label{eq:free_action_difference_quotient}
\begin{eqnarray}
\label{eq:free_action_difference_quotient_right}
\nonumber \hspace{-3em} && S_M^0 (\bar{\varphi}, \varphi)\\
\hspace{-3em} &=& \sum_\alpha \! \Delta \! \sum_{m = 1}^M \! \bar{\varphi}^m_\alpha \! \bigg[ \ii \frac{\varphi^m_\alpha \! - \! \varphi^{m - 1}_\alpha}{\Delta} - (1 \! - \! \ii \eta) \varepsilon_\alpha \varphi^{m - 1}_\alpha \! \bigg] \bigg|_\mathcal{B}\\
\label{eq:free_action_difference_quotient_left}
\hspace{-3em} &=& \sum_\alpha \! \Delta \! \sum_{m = 0}^{M - 1} \! \bigg[ \ii \frac{\bar{\varphi}^m_\alpha \! - \! \bar{\varphi}^{m + 1}_\alpha}{\Delta} - (1 \! - \! \ii \eta) \varepsilon_\alpha \bar{\varphi}^{m + 1}_\alpha \! \bigg] \varphi^m_\alpha \bigg|_\mathcal{B} \! .
\end{eqnarray}
\end{subequations}
In Eq.~(\ref{eq:free_action_difference_quotient}) some addends were introduced artificially; they evaluate to zero due to the extended boundary conditions
\begin{equation}
\mathcal{B} = \big\{ \bar{\varphi}^0_{\alpha \le N} = 0, \, \bar{\varphi}^M_{\alpha > N} = 0, \, \varphi^M_{\alpha \le N} = 0, \, \varphi^0_{\alpha > N} = 0 \big\}.
\end{equation}
In the limit $M \to \infty$, the free action can thus be written as
\begin{equation}
S^0 [\bar{\varphi}, \varphi] = \sum_{\alpha' \alpha} \int_{- t_0}^{t_0} \dd t' \int_{- t_0}^{t_0} \dd t \bar{\varphi}_{\alpha'} (t') Q_{\alpha' \alpha} (t', t) \varphi_\alpha (t) \bigg|_\mathcal{B}
\end{equation}
with the boundary conditions in continuous form
\begin{eqnarray}
\label{eq:extended_boundary_conditions_continuous}
\nonumber \mathcal{B} &=& \big\{ \bar{\varphi}_{\alpha \le N} (- t_0) = 0, \, \bar{\varphi}_{\alpha > N} (t_0) = 0,\\
&& \hphantom{\big\{} \varphi_{\alpha \le N} (t_0) = 0, \, \varphi_{\alpha > N} (- t_0) = 0 \big\}
\end{eqnarray}
and with the differential operator
\begin{subequations}
\label{eq:inverse_free_propagator_continuous}
\begin{equation}
Q_{\alpha' \alpha} (t', t) = \delta_{\alpha' \alpha} \delta (t' - t) \big[ \ii \partial_t - (1 - \ii \eta) \varepsilon_\alpha \big]
\end{equation}
or
\begin{equation}
Q_{\alpha' \alpha} (t', t) = \delta_{\alpha' \alpha} \Big[ - \ii \overset{\leftarrow}{\partial}_{\! t'} - (1 - \ii \eta) \varepsilon_\alpha \Big] \delta (t' - t)
\end{equation}
\end{subequations}
corresponding to the expression (\ref{eq:free_action_difference_quotient_right}) or (\ref{eq:free_action_difference_quotient_left}), respectively.
In the latter version, the time derivative acts to the left.
Indeed, the continuous form of the free propagator (\ref{eq:free_propagator_continuous}) is the inverse, i.e., the Green function, of $Q_{\alpha' \alpha} (t', t)$.
It is unique in satisfying the boundary conditions
\begin{align}
\nonumber g_{\alpha \le N} (t, - t_0) &= 0, & g_{\alpha > N} (t, t_0) &= 0,\\
g_{\alpha \le N} (t_0, t') &= 0, & g_{\alpha > N} (- t_0, t') &= 0,
\end{align}
which follow from Eq.~(\ref{eq:extended_boundary_conditions_continuous}).

In the continuum limit, the interaction part of the action is
\begin{equation}
\label{eq:interaction_part_of_action_continuous}
S^\text{int} [\bar{\varphi}, \varphi] = - (1 - \ii \eta) \int_{- t_0}^{t_0} \dd t H_\text{int} \big( \bar{\varphi} (t), \varphi (t^-) \big) \bigg|_\mathcal{B}
\end{equation}
and the source part of the action is
\begin{eqnarray}
\nonumber && S^\text{source} [\bar{\varphi}, \varphi; \bar{J}, J]\\
&=& - \sum_\alpha \int_{- t_0}^{t_0} \dd t \big[ \bar{J}_\alpha (t) \varphi_\alpha (t) + \bar{\varphi}_\alpha (t) J_\alpha (t) \big] \bigg|_\mathcal{B}.
\end{eqnarray}

\section{Diagrammatic expansion}
\label{sec:diagrammatic_expansion}

In this appendix we briefly describe the steps that lead to the diagrammatic expansion of Green functions.
Given the functional-integral representation (\ref{eq:generating_functional_continuous}) of the generating functional in the interacting case, we can follow the standard procedure as outlined in Chap.~2 of Ref.~\cite{Negele88} for the case of the Matsubara formalism.
Note that Chap.~3 of Ref.~\cite{Negele88} on the zero-temperature formalism does not provide an expression corresponding to our Eq.~(\ref{eq:generating_functional_continuous}).

We expand the integrands in Eq.~(\ref{eq:generating_functional_continuous}) in powers of the interaction.
The interacting Green functions are thereby expressed in terms of noninteracting Green functions.
The latter can be calculated with the Wick theorem that results when the functional derivatives in Eq.~(\ref{eq:Green_functions_from_generating_functional}) are applied to the noninteracting generating functional, which is in the continuous notation of Sec.~\ref{sec:continuous_notation} given by
\begin{equation}
\mathcal{G}_\eta^0 [\bar{J}, J] = e^{- \ii \bar{J} g J}.
\end{equation}
By the standard steps, the interacting Green functions $G_\eta$ can be represented by sums of diagrams made out of external points, interaction vertices, and free-propagator lines.
Due to the denominator in Eq.~(\ref{eq:generating_functional_continuous}), all clusters that are not linked to the external points cancel out from the diagrams.
We can obtain an efficient representation by employing unlabeled Hugenholtz vertices, see Chap.~2 of Ref.~\cite{Negele88}.
The value of a specific diagram is then given by
\begin{equation}
\label{eq:value_of_general_diagram}
\frac{(-1)^P (-1)^{n_\text{loop}}}{2^{n_\text{eq}} S} \left( \prod \ii \bar{v} \right) \left( \prod g \right)
\end{equation}
with implicit contractions of all internal multi-indices.
$(-1)^P$ is the sign of the permutation which is given by the external indices $x_i'$ and $x_{P(i)}$ being connected, $n_\text{loop}$ is the number of internal closed loops, $n_\text{eq}$ is the number of pairs of equivalent lines, and $S$ is the diagram symmetry factor, see Chap.~2 of Ref.~\cite{Negele88}.

\section{Frequency representation}
\label{sec:frequency_representation}

Here, we describe the Fourier transformation used to go from time arguments in the diagrammatic expansion over to frequency arguments.
Since the Green functions (\ref{eq:definition_of_Green_functions}), the free propagator (\ref{eq:free_propagator_continuous}), and the bare vertex (\ref{eq:bare_vertex_general}) are time-translationally invariant, it is advantageous to work in frequency representation, where each of these quantities becomes frequency conserving and where convolutions of time that occur in the diagrams are transformed into simple products.
We note that the limit $t_0 \to \infty$ has to be performed first so that the boundaries of the Fourier integrals $\int_{- \infty}^\infty \dd t \, e^{\pm \ii \omega t} \ldots$ can be infinite.

The frequency-dependent free propagator is defined as
\begin{equation}
g_{\alpha \alpha'} (\omega, \omega') = \int_{- \infty}^\infty \dd t \int_{- \infty}^\infty \dd t' e^{\ii (\omega t - \omega' t')} g_{\alpha \alpha'} (t, t').
\end{equation}
The integrals converge thanks to the positive infinitesimal $\eta$ that was introduced in the context of the generating functional, see Eq.~(\ref{eq:time_evolution_operator}).
The result is
\begin{subequations}
\begin{eqnarray}
g_{\alpha \alpha'} (\omega, \omega') &=& 2 \pi \delta (\omega - \omega') \delta_{\alpha \alpha'} g_\alpha (\omega)\\
\label{eq:frequency_representation_of_free_propagator_diagonal} g_\alpha (\omega) &=& \frac{e^{\ii \omega \eta'}}{\omega - \varepsilon_\alpha + \ii \eta \sgn \varepsilon_\alpha},
\end{eqnarray}
\end{subequations}
where we made the replacement $\eta \varepsilon_\alpha \to \eta \sgn \varepsilon_\alpha$.
The positive infinitesimal $\eta'$ is necessary to obtain the correct equal-time value $g_{\alpha \alpha'} (t, t)$ from the inverse Fourier transformation.
It is relevant only for Hartree-Fock-type propagator loops.
The bare vertex is Fourier transformed into
\begin{eqnarray}
\nonumber && \bar{v}_{\alpha'_1 \alpha'_2 \alpha_1 \alpha_2} (\omega'_1, \omega'_2, \omega_1, \omega_2)\\
&=& 2 \pi \delta (\omega'_1 + \omega'_2 - \omega_1 - \omega_2) (1 - \ii \eta) \bar{v}_{\alpha'_1 \alpha'_2 \alpha_1 \alpha_2}.
\end{eqnarray}

The frequency-dependent interacting Green functions
\begin{eqnarray}
\nonumber && G(\omega_1, \ldots| \ldots, \omega'_n)\\
\nonumber &=& \int_{- \infty}^\infty \dd t_1 \ldots \int_{- \infty}^\infty \dd t'_n e^{\ii (\omega_1 t_1 + \cdots - \omega'_n t'_n)} G(t_1, \ldots| \ldots, t'_n)\\
& \propto & \delta (\omega_1 + \cdots + \omega_n - \omega'_1 - \cdots - \omega'_n)
\end{eqnarray}
are well-defined whenever long-term correlations decay so that the Fourier integrals converge.
They can be calculated by applying the diagrammatic rules from appendix \ref{sec:diagrammatic_expansion} to obtain $G_\eta$ and finally taking the limit $G = \lim_{\eta \to 0^+} G_\eta$.

\section{Derivation of 1PI flow equations}
\label{sec:derivation_of_flow_equations}

In the following we outline how to arrive at the generating functional for the 1PI vertex functions, starting from $\mathcal{G}_\eta [\bar{J}, J]$.
Based on these steps, we then present a concise derivation of the FRG flow equations of the 1PI vertex functions.
It adapts the established approach (see, e.g., Ref.~\cite{Negele88, Metzner12}) to the zero-temperature formalism.
We will use matrix notation with multi-indices $x = (\alpha, t)$ or $x = (\alpha, \omega)$, depending on the chosen representation.

Let us define
\begin{equation}
\mathcal{W}_\eta [\bar{J}, J] = \ln \mathcal{G}_\eta [\bar{J}, J] = \lim_{t_0 \to \infty} \ln \frac{Z_\eta [\bar{J}, J]}{Z_\eta [0, 0]},
\end{equation}
which can be shown to be the generating functional of the connected Green functions via the replica technique.
Next, we define the so-called effective action as the Legendre transform of $\mathcal{W}_\eta [\bar{J}, J]$,
\begin{equation}
\mathcal{U}_\eta [\bar{\psi}, \psi] = - \bar{J} \psi + \bar{\psi} J + \mathcal{W}_\eta [\bar{J}, J],
\end{equation}
where $\bar{J} = \bar{J} [\bar{\psi}, \psi]$ and $J = J [\bar{\psi}, \psi]$ are the inverse relations to the definitions of the new variables
\begin{eqnarray}
\bar{\psi}_{x'} [\bar{J}, J] = \frac{\delta \mathcal{W}_\eta}{\delta J_{x'}}, \qquad \psi_x [\bar{J}, J] = \frac{\delta \mathcal{W}_\eta}{\delta \bar{J}_x}.
\end{eqnarray}
Note that this (standard) notation is misleading as $\bar{\psi}$ and $\psi$ are not conjugated to each other even if $\bar{J}$ and $J$ are.
Lastly, we define the generating functional of the 1PI vertex functions
\begin{equation}
\Gamma_\eta [\bar{\psi}, \psi] = \mathcal{U}_\eta [\bar{\psi}, \psi] - \ii \bar{\psi} Q \psi.
\end{equation}
The 1PI vertex functions are generated from it via
\begin{eqnarray}
&& \gamma (x'_1, \ldots, x'_n| x_1, \ldots, x_n)\\
\nonumber &=& \lim_{\eta \to 0^+} (-1)^n \ii \frac{\delta^{2n} \Gamma_\eta [\bar{\psi}, \psi]}{\delta \bar{\psi}_{x'_1} \ldots \delta \bar{\psi}_{x'_n} \delta \psi_{x_n} \ldots \delta \psi_{x_1}} \bigg|_{\bar{\psi} = 0 = \psi} .
\end{eqnarray}
Diagrammatically, they are the sums of all one-particle irreducible diagrams, which cannot be disconnected by removing any single propagator line and which are evaluated with an additional prefactor $\ii^{1 - n}$ compared to Eq.~(\ref{eq:value_of_general_diagram}).

In the 1PI FRG formalism, the problem of determining the 1PI vertex functions is recast into the task to solve a set of differential equations that map the 1PI vertex functions for an easily solvable system to the ones for the system of interest.
To this end one introduces a flow parameter $\lambda$ into the free propagator $g \to g_\lambda$.
Then also the functional $Z_\eta [\bar{J}, J]$ acquires a $\lambda$-dependence through the inverse free propagator $Q \to Q_\lambda$.
Its flow equation reads
\begin{equation}
\label{eq:flow_of_Z}
\dot{Z} = \ii \frac{\delta}{\delta J} \dot{Q} \frac{\delta}{\delta \bar{J}} Z.
\end{equation}
For conciseness we have suppressed all dependencies, used the shorthand notation $\dot{Z} = \dd Z / \dd \lambda$, and considered $\delta / \delta J$ as a row vector and $\delta / \delta \bar{J}$ as a column vector.
Since the definition of $Z_\eta [\bar{J}, J]$ does not contain the limit $t_0 \to \infty$ as it would otherwise diverge, see Eq.~(\ref{eq:Gaussian_integral}), the time integrations in Eq.~(\ref{eq:flow_of_Z}) are restricted to the finite interval $[- t_0, t_0]$.
This is not the case for the other equations in this appendix.
For $\mathcal{W}_\eta [\bar{J}, J]$ one obtains the flow equation
\begin{equation}
\dot{\mathcal{W}} = \ii \left( \frac{\delta \mathcal{W}}{\delta J} \dot{Q} \frac{\delta \mathcal{W}}{\delta \bar{J}} - \trace \dot{Q} \frac{\delta^2 \mathcal{W}}{\delta \bar{J} \delta J} \right) - \trace \dot{Q} G,
\end{equation}
where $G$ stands for the single-particle Green function and the second derivatives $\delta^2 \mathcal{W} / \delta \bar{J}_x \delta J_{x'}$ form a matrix with row index $x$ and column index $x'$.
This in turn leads to
\begin{equation}
\dot{\Gamma} = - \ii \trace \dot{Q} (u^{-1})_{++} - \trace \dot{Q} G
\end{equation}
with the matrix
\begin{equation}
u_{X X'} = \frac{\delta^2 \mathcal{U}}{\delta \psi_X \delta \psi_{\bar{X}'}},
\end{equation}
for which the multi-index was extended to $X = (c, x)$ with $c = \pm$ and we defined $\bar{X} = (- c, x)$, $\psi_{(+, x)} = \bar{\psi}_x$, and $\psi_{(-, x)} = \psi_x$.
With this notation the resulting flow equations of the 1PI vertex functions can be written as
\begin{eqnarray}
\label{eq:general_flow_of_1PI_vertex_functions}
&& \dot{\gamma} (x'_1, \ldots, x'_n| x_1, \ldots, x_n)\\
\nonumber &=& \lim_{\eta \to 0^+} (-1)^n \trace \dot{Q} \Bigg( \frac{\delta^{2 n} u^{-1}}{\delta \bar{\psi}_{x'_1} \ldots \delta \psi_{x_1}} \bigg|_{\bar{\psi} = 0 = \psi} \Bigg)_{\! \! ++}.
\end{eqnarray}
The right-hand side of each of these flow equations corresponds to a sum over ring diagrams composed of 1PI $m$-particle vertex functions with $m = 2, \ldots, n + 1$ that are connected by full single-particle Green functions and one single-scale propagator
\begin{equation}
\label{eq:single-scale_propagator_general}
S = - G \dot{Q} G = G g^{-1} \dot{g} g^{-1} G.
\end{equation}

\section{Generalization of the zero-temperature formalism}
\label{sec:generalization_of_zero-temperature_formalism}

In spite of its name, the zero-temperature formalism can in certain cases be used to compute the properties of systems in particular excited states.
Consider an $N$-particle eigenstate of $H_{0}$ given by
\begin{equation}
\big| \tilde{\Phi} \big\rangle = a_{\tilde{1}}^\dagger \ldots a_{\tilde{N}}^\dagger | 0 \rangle,
\end{equation}
where $\tilde{1}, \ldots, \tilde{N}$ denote some single-particle eigenstates of $H_0$.
We do not require that $\{ \tilde{1}, \ldots, \tilde{N} \} = \{ 1, \ldots, N \}$, i.e., $\big| \tilde{\Phi} \big\rangle$ does not have to be the ground state of $H_0$.
Let $\big| \tilde{\Psi} \big\rangle$ denote the lowest-energy eigenstate of $H$ that has a nonzero overlap with $\big| \tilde{\Phi} \big\rangle$.
We assume $\big| \tilde{\Psi} \big\rangle$ to be unique.
If $\big| \tilde{\Phi} \big\rangle$ is orthogonal to the ground state of $H$, then $\big| \tilde{\Psi} \big\rangle $ is not that ground state.
$\big| \tilde{\Phi} \big\rangle$ evolves into $\big| \tilde{\Psi} \big\rangle$ under a damped but suitably normalized time evolution.
To be more precise, we define
\begin{equation}
\tilde{Z}_\eta [\bar{J}, J] = \big\langle \tilde{\Phi} \big| U^{(\eta)}_{\bar{J}, J} (t_0, - t_0) \big| \tilde{\Phi} \big\rangle
\end{equation}
and
\begin{equation}
\tilde{\mathcal{G}}_\eta [\bar{J}, J] = \lim_{t_0 \to \infty} \frac{\tilde{Z}_\eta [\bar{J}, J]}{\tilde{Z}_\eta [0,0]}
\end{equation}
on the analogy of Eq.~(\ref{eq:generating_functional_definition}) to (\ref{eq:time_evolution_operator}).
Then $\tilde{\mathcal{G}}_\eta [\bar{J}, J]$ generates the Green functions of the system in the state $\big| \tilde{\Psi} \big\rangle$,
\begin{eqnarray}
\label{eq:generalized_definition_of_Green_functions}
\nonumber && \tilde{G} (\alpha_1 t_1, \ldots| \ldots, \alpha'_n t'_n)\\
&=& (- i)^n \left\langle \tilde{\Psi} \middle| \mathcal{T} a_{\alpha_1} (t_1) \ldots a_{\alpha'_1}^\dagger (t'_1) \middle| \tilde{\Psi} \right\rangle.
\end{eqnarray}

The steps done in Sec.~\ref{sec:definition_of_Green_functions_and_their_generating_functional} to \ref{sec:continuous_notation} can be applied also to this situation.
Instead of $\alpha \le N$ and $\alpha > N$, one now distinguishes between $\alpha \in \{ \tilde{1}, \ldots, \tilde{N} \}$ and $\alpha \notin \{ \tilde{1}, \ldots, \tilde{N} \}$, referring to levels which are occupied and empty in the state $\big| \tilde{\Phi} \big\rangle$, respectively.
However, it is not evident that a diagrammatic expansion analogous to appendix \ref{sec:diagrammatic_expansion} is well-defined.
If a single-particle state $\alpha$ with $\varepsilon_\alpha < 0$ is unoccupied in $\big| \tilde{\Phi} \big\rangle$, then the corresponding propagator $\tilde{g}_\alpha (t) = - \ii e^{- (\ii + \eta) \varepsilon_\alpha t} \Theta (t - 0^+)$ diverges exponentially with time.
The same holds for single-particle states with positive eigenenergy that are occupied in $\big| \tilde{\Phi} \big\rangle$.
(We note that if $\big| \tilde{\Phi} \big\rangle$ is not the ground state of $H_0$, one cannot avoid all such divergences by simply shifting the reference point for single-particle energies.)
Therefore, the convergence of time integrations that arise in diagrams has to be checked.
In the following we present simple cases in which a diagrammatic expansion is possible, focusing on the treatment of absorption and emission of X-rays in metals.

We start by considering the case that $\big| \tilde{\Phi} \big\rangle = \big| \tilde{\Psi} \big\rangle$ is a common eigenstate of $H_0$ and $H$.
Then the damping factor $\eta$ and the limit $t_0 \to \infty$ are not required in the definition of $\tilde{Z} [\bar{J}, J]$ and $\tilde{\mathcal{G}} [\bar{J}, J]$ to generate the Green function (\ref{eq:generalized_definition_of_Green_functions}). Instead, it suffices to choose $t_0 \ge \max \{ |t_1|, \ldots, |t'_n| \}$.
Thus, the time integrations that arise in diagrams are restricted to the finite interval $[- t_0, t_0]$ and converge.
In order to derive a frequency representation, one takes the limit $t_0 \to \infty$ and only now introduces appropriate dampings $\eta$ into the retarded and advanced free propagators such that the Fourier integrals converge.
The result for the free propagator is as in appendix \ref{sec:frequency_representation},
\begin{equation}
\label{eq:generalized_free_propagator}
\tilde{g}_\alpha (\omega) = e^{\ii \omega \eta'} \! \begin{cases} \! (\omega - \varepsilon_\alpha + \ii \eta)^{-1}, & \alpha \notin \{ \tilde{1}, \ldots, \tilde{N} \}\\ \! (\omega - \varepsilon_\alpha - \ii \eta)^{-1}, & \alpha \in \{ \tilde{1}, \ldots, \tilde{N} \}. \end{cases}
\end{equation}

We present two examples of this situation in the context of the model for X-ray absorption in metals with
\begin{subequations}
\begin{eqnarray}
H_0 &=& \sum_k \varepsilon_k a_k^\dagger a_k + \varepsilon_d a_d^\dagger a_d,\\
H_\text{int} &=& - \frac{U}{V} \sum_{kk'} a_{k'}^\dagger a_k a_d a_d^\dagger\\
&=& - \frac{U}{V} \sum_{kk'} a_{k'}^\dagger a_k + \frac{U}{V} \sum_{kk'} a_{k'}^\dagger a_d^\dagger a_d a_k\\
&=& H_\text{int}^{(1)} + H_\text{int}^{(2)}.
\end{eqnarray}
\end{subequations}
First, we define $\big| \tilde{\Phi} \big\rangle$ as the state in which the deep level as well as all plane-wave states $k$ in the lower half of the conduction band are occupied, while the upper half of the conduction band is empty.
For $\varepsilon_d < 0$ this state is the ground state of $H_0$, but for $\varepsilon_d > 0$ it is an excited eigenstate of $H_0$.
Furthermore, for sufficiently negative $\varepsilon_d$ the state $\big| \tilde{\Phi} \big\rangle$ is the ground state of $H$; otherwise, it is an excited eigenstate of $H$.
In any case, the formalism allows to compute the Green functions of the system in this state.
As a consequence the function $\chi(\nu)$ depends smoothly on $\varepsilon_d$ as indicated in Eq.~(\ref{eq:zero-temperature_Ward_identity}) so that the transition to $\tilde{\varepsilon}_d = 0$ in Sec.~\ref{sec:setting_renormalized_deep_level_to_zero} is possible.

For the second example, we consider the system with an empty deep level.
The single-particle eigenstates of $H'_0 = H_0 + H_\text{int}^{(1)}$ consist of the deep state, a conduction band of scattering states, and one bound state which originates from the attractive potential of the deep hole and which has an energy below the conduction band.
We define $\big| \tilde{\Phi}' \big\rangle$ as the state with the deep level empty but the bound state and the $(N-1)$ lowest scattering states occupied.
This state is a common eigenstate of $H'_0$ and $H$.
Again, it depends on the value of $\varepsilon_d$ whether it is the respective ground state or an excited state.
In each instance the formalism can be used to compute the Green functions in the state $\big| \tilde{\Phi}' \big\rangle$.
In the corresponding perturbative expansion, $H_\text{int}^{(2)}$ is treated as perturbation to $H'_0$.
The free propagator in frequency representation is analogous to Eq.~(\ref{eq:generalized_free_propagator}),
\begin{equation}
\label{eq:generalized_free_propagator_including_single-particle_interaction}
\tilde{g}'_{\alpha'} (\omega) = e^{\ii \omega \eta'} \! \begin{cases} \! (\omega - \varepsilon'_{\alpha'} + \ii \eta)^{-1}, & \! \alpha' \! \text{ unoccupied in } \! \big| \tilde{\Phi}' \big\rangle\\ \! (\omega - \varepsilon'_{\alpha'} - \ii \eta)^{-1}, & \! \alpha' \! \text{ occupied in } \! \big| \tilde{\Phi}' \big\rangle. \end{cases}
\end{equation}
Here, $\alpha'$ denotes a single-particle eigenstate of $H'_0$ with energy $\varepsilon'_{\alpha'}$.
This approach can be used to study the rate of stimulated X-ray emission, cf. Ref.~\cite{Roulet69}.

As a final, related case, we present an alternative way to access the Green functions in the state $\big| \tilde{\Phi}' \big\rangle$ from above and thereby treat X-ray emission, cf. Ref.~\cite{Roulet69}.
For this we consider the state $\big| \tilde{\Phi}_1 \big\rangle$ with the deep level empty and the lowest $N$ plane-wave conduction states (not scattering states) occupied.
This is an eigenstate of $H_0$.
Under a damped but suitably normalized time evolution, it evolves into $\big| \tilde{\Phi}' \big\rangle$.
If we define $\tilde{Z}_\eta [\bar{J}, J]$ as the expectation value of $U^{(\eta)}_{\bar{J}, J} (t_0, - t_0)$ in the state $\big| \tilde{\Phi}_1 \big\rangle$, the corresponding functional $\tilde{\mathcal{G}}_\eta [\bar{J}, J]$, involving the limit $t_0 \to \infty$, generates the Green functions of the interacting system in its stationary state $\big| \tilde{\Phi}' \big\rangle$.
This leads to an expansion in $H_\text{int}^{(1)} + H_\text{int}^{(2)}$ as perturbation to $H_0$.
Although the retarded free deep-state propagator $\tilde{g}_d (t) = - \ii e^{- (\ii + \eta) \varepsilon_d t} \Theta (t - 0^+)$ diverges exponentially for $t \to \infty$ if $\varepsilon_d < 0$, all time integrations arising in diagrams converge.
The reason is as follows.
Since each two-particle vertex has both an incoming and an outgoing deep-state leg and since time strictly increases in direction of the deep-state lines, the time arguments of the two-particle vertices are restricted to the finite interval spanned by the external time arguments of the Green function.
Therefore, integrations with respect to the time arguments of two-particle vertices converge.
Only the single-particle vertices produced by $H_\text{int}^{(1)}$ can appear on the whole time axis.
They have an incoming and an outgoing plane-wave line only.
Since the occupied plane-wave levels in $\big| \tilde{\Phi}_1 \big\rangle$ are the lowest ones, we can choose the reference point for single-particle energies to lie between the occupied and unoccupied plane-wave levels.
Then the dampings $\eta \varepsilon_k$ contained in the free plane-wave propagators ensure that they are exponentially suppressed for $t \to \pm \infty$.
We note that while this particular choice of the reference point makes it easy to realize that the time integrals converge also for the single-particle vertices, it is actually not necessary for the convergence.
In fact, only differences of plane-wave levels enter the integrals with respect to times of single-particle vertices.
Since in all situations the time integrations ultimately converge, it is possible to adjust the dampings in the individual free propagators $\tilde{g}_\alpha (t)$ such that all retarded ones are suppressed for $t \to \infty$ and all advanced ones are suppressed for $t \to - \infty$.
Diagrams can then be evaluated as well in Fourier space with
\begin{equation}
\tilde{g}_\alpha (\omega) = e^{\ii \omega \eta'} \! \begin{cases} \! (\omega - \varepsilon_\alpha + \ii \eta)^{-1}, & \! \alpha \! \text{ unoccupied in } \! \big| \tilde{\Phi}_1 \big\rangle\\ \! (\omega - \varepsilon_\alpha - \ii \eta)^{-1}, & \! \alpha \! \text{ occupied in } \! \big| \tilde{\Phi}_1 \big\rangle. \end{cases}
\end{equation}
We expect that dressing this free propagator, which corresponds to $H_0$, with the single-particle vertices from $H_\text{int}^{(1)}$ leads to the propagator (\ref{eq:generalized_free_propagator_including_single-particle_interaction}), which corresponds to $H'_0$.

\bibliography{parquet_approx_one-loop_RG_equiv_leading-log}

\end{document}